\documentclass[journal,twoside,final]{IEEEtran}
\usepackage{amssymb, amsmath, amsthm, bm, tikz, cite,makecell}
\usetikzlibrary{positioning}

\ifCLASSOPTIONcompsoc
\usepackage[caption=false,font=normalsize,labelfont=sf,textfont=sf]{subfig}
\else
\usepackage[caption=false,font=footnotesize]{subfig}
\fi

\newcommand{\shot}[1]{n_\text{sh}\left(#1\right)}
\newcommand{\thermal}[1]{n_\text{th}\left(#1\right)}
\newcommand{\sshot}{\sigma_\text{sh}}
\newcommand{\sthermal}{\sigma_\text{th}}
\newcommand{\relates}{\equiv}
\newcommand{\nrelates}{\not\equiv}
\newcommand{\expect}[1]{\mathbb{E}\left[#1\right]}
\newcommand{\var}[1]{\text{Var}\left(#1\right)}
\newcommand{\cov}[2]{\text{Cov}\left(#1,#2\right)}
\newcommand{\subreference}[2]{\ref{#1}\protect\subref{#2}}

\newtheorem{theorem}{Theorem}
\newtheorem{example}{Example}

\title{Direct Detection Under Tukey Signalling}
\author{Amir Tasbihi~\IEEEmembership{Graduate Student
Member, IEEE} and Frank R.
Kschischang~\IEEEmembership{Fellow, IEEE}\thanks{Submitted on May 26th, 2021; revised July 11th and August 7th, 2021.  The authors
are with the Edward S. Rogers Sr.\ Dept.\ of Electrical \& 
Computer Engineering, University of Toronto, Toronto, ON M5S
3G4, Canada.  Email:
\texttt{\{tasbihi,frank\}@ece.utoronto.ca.}}}
\begin{document}
\maketitle
\begin{abstract}
A new direct-detection-compatible signalling scheme is proposed
for fiber-optic communication over short distances.
Controlled inter-symbol interference 
is exploited to
extract phase information, thereby achieving
spectral efficiencies about one bit less, per second per hertz, 
of those of a coherent detector.
\end{abstract}
\begin{section}{Introduction}
\label{sec:introduction}
\IEEEPARstart{D}{irect} detection or, synonymously, square-law detection,
is a nonlinear waveform detection scheme
based upon measuring the squared magnitude of 
a complex-valued waveform. It appears in various scientific fields, 
\textit{e.g.},
crystallography~\cite{crystallography},
radio astronomy~\cite{astronomy1,astronomy2},
biomedical spectroscopy~\cite{bio1},
detection and estimation theory~\cite{radar1,radar2,radar3}, etc.
This paper deals with the application of square-law detection 
in fiber-optic communication systems\cite{agrawal,kumar},
particularly those with short transmission  
length, \textit{e.g.}, less than $10$~km.
Such systems occur, \textit{e.g.}, in
rack-to-rack data transmission within data centers. 

Since square-law detectors
base their decision only on the magnitude of the received 
complex-valued waveform---unlike coherent detectors which also access phase
information---one might get the intuitive impression
that the information rate of a communication channel 
under square-law detection should be 
roughly half the information rate of the same channel
under coherent detection~\cite{wrong_capacity1,wrong_capacity2,
		wrong_capacity3,wrong_capacity4}. 
This intuition is actually correct for
\textit{intensity modulation with direct detection} (IM/DD)
systems, which modulate only the 
magnitude of the transmitted waveform. 
In particular, the transmitted symbol in IM/DD systems belong
to a finite real set whose elements have distinct intensity, \textit{i.e.},
squared magnitude.    
Due to their 
simple transceiver structure, IM/DD systems
have been extensively used in short-reach optical communications and 
have been investigated deeply in the literature~\cite{imdd0,imdd0a,imdd1,imdd2,
			imdd3,imdd4,imdd5}.   
However, the simplicity of IM/DD systems comes at the price 
of losing about half the degrees of freedom
compared with coherent detection, which enables detection in
the complex field.

Somewhat counter-intuitively, 
it can be shown that the capacity
of waveform channels under square-law detection of
bandlimited~\cite{capacity_bandlimited}
and time-limited~\cite{capacity_timelimited} signals is, in fact, at most
one bit less, per second per hertz, than the capacity under
coherent detection!
Clearly, this is possible when the transmitted symbols may assume
complex values. 
Thus, the price to pay for the convenience of direct detection may not be
as high as intuition would suggest.
Unfortunately, \cite{capacity_bandlimited} and 
\cite{capacity_timelimited} do not present a practical scheme to achieve
the lower bound on the capacity derived in those papers. 

Due to the simplicity of the receiver's optical front-end in direct detection, 
finding efficient communication schemes compatible with square-law detection
is an active area of research.
A popular recent scheme is the 
\textit{Kramers--Kronig receiver} \cite{kk1}, which has been investigated
thoroughly in the optical communication literature in the last few years.
The Kramers--Kronig receiver enables recovery of a 
complex-valued waveform from its squared magnitude,
which in turn enables the use of digital signal processors to
mitigate dispersion, making direct detection
viable for longer transmission lengths ($>100~$km) than
previously thought,
where dispersion is a hindering factor~\cite{kk2,kk3,kk6}.    
These merits come with drawbacks however:
a high required carrier-to-signal power ratio~\cite{kk7} and 
increased sampling rates necessitated by
spectrum-broadening operations performed after the square-law device. 

In this paper, a new 
direct-detection-compatible data transmission
scheme is proposed that exploits deliberately-introduced
inter-symbol interference (ISI) to extract phase information.
The deliberate introduction of ISI in digital communication
via so-called \emph{partial-response coding}~\cite{pr1} dates
back to the early 1960s, and allows for the design of line codes
with prescribed spectral nulls and other
useful properties~\cite{pr2,pr3,pr4,pr5,pr6,pr7}.
Partial-response
systems arise in high-density
magnetic recording systems~\cite{mr1,mr2,mr3} where the read-channel
is equalized to achieve a particular response characteristic.
Deliberate intersymbol interference also
arises in faster-than-Nyquist signalling schemes
\cite{ftn1,ftn2,ftn3,ftn4} which have been proposed
for wireless and optical communications.
In this paper we propose a new application of controlled ISI: namely,
to extract phase information.

As a toy example to illustrate that ISI can be beneficial, let
$z_1$ and $z_2$ be complex numbers. Then, from  
$|z_1|^2$, $|z_2|^2$, and $|z_1+z_2|^2$ (an ISI term), 
one can retrieve the 
phase difference between $z_1$ and $z_2$, up to a sign ambiguity.

As another example, let 
\[
g(t)=\sum_{\ell=0}^{m}g_\ell \text{~sinc}(t-\ell),
\]
where $g_0,\ldots,g_m\in\mathbb{C}$ and
$\text{sinc}(t)\triangleq\frac{\sin(\pi t)}{\pi t}$. 
Here $g(t)$ represents the signal transmitted in an ideal bandlimited
quadrature amplitude modulation (QAM) data transmission system with
unit symbol rate.
Note that $|g(t)|^2 = g(t) \cdot g^*(t)$, being the product
of bandlimited signals,
has twice the bandwidth of $g(t)$, and thus there is a possibility
to recover
$g_0,\ldots,g_m$ from samples of $|g(t)|^2$
at $t=\frac{k}{2}$, where $k=0,\ldots,2m$ (i.e., samples
taken at twice the symbol rate).
Since
\begin{equation*}
\left|g\left(\frac{k}{2}\right)\right|^2=
\begin{cases}
\left|g_\frac{k}{2}\right|^2, & \text{if $k$ is even;}\\
\left|\sum_{\ell=0}^{m}g_\ell\text{~sinc}\left(\frac{k}{2}-\ell\right)\right|^2,
&\text{if $k$ is odd,}
\end{cases}
\end{equation*}
one can easily recover
the magnitudes of $g_0,\ldots,g_m$ 
from the samples with an even $k$, while their
phases are embedded in samples with an odd $k$. 
Unfortunately, since all 
$g_\ell$'s contribute to the samples at half-integer times, recovering 
phase information is an intractable problem, 
even for moderately small values of $m$.
Therefore, while ISI is useful to extract phase information, 
an excessive ISI would demand complex processing. This is the rationale for 
claiming that ``controlled ISI'' is needed. 

The rest of the paper is organized as follows. The system model, including the
transmitter, the channel, and the receiver, is described in
Sec.~\ref{sec:system_model}. In particular, the new signalling scheme is
proposed in Sec.~\ref{subsec:tukey_signalling}.  The scheme is validated via
numerical simulations, whose results are given in
Sec.~\ref{sec:numerical_simulation}.  The implementation
complexity of the proposed scheme (in terms of digital-to-analog
and analog-to-digital conversions per symbol) is
compared with those of a coherent detector and a Kramers--Kronig detector in
Sec.~\ref{sec:complexity}.
Finally, concluding remarks are provided in
Sec.~\ref{sec:discussion}.  

To simplify the discussion throughout this
paper,
for our
proposed system and for all systems with which we compare, we consider
transmission of information on a single polarization only.
The extension to polarization-multiplexed transmissions is,
at least conceptually, straightforward.

Throughout this paper, vectors are denoted by lower-case
bold letters, \textit{e.g.}, $\bm{v}$.
For a vector $\bm{v}$ of length $m$,
$\bm{v}[k]$ denotes its $k^\text{th}$ entry,
where $k\in\{0,\ldots,m-1\}$.
The cardinality of a finite set 
$\mathcal{A}$ is denoted by $|\mathcal{A}|$.
The expected value and the variance of a random 
variable $X$ are denoted as $\expect{X}$ and $\var{X}$, respectively. 
Likelihood functions will always be denoted as $f$, with
arguments chosen to indicate the random variables involved;
for example,
the conditional probability density function of a random
variable $Y$ at a point $y \in \mathbb{R}$ given that
a random variable $X$ takes value $x$ is denoted simply as
$f(y\mid x)$ rather than the more cumbersome $f_{Y \mid X}(y \mid x)$.
The notation
$X\sim\mathcal{N}(\mu,\sigma^2)$ indicates that random variable
$X$ has a Gaussian distribution with mean $\mu$ and variance $\sigma^2$.
Finally, the positive real numbers are denoted as $\mathbb{R}^{>0}$. 
\end{section}

\begin{section}{The System Model}
\label{sec:system_model}
In this section we describe the system model, shown in 
Fig.~\ref{fig:system_model}. For simplicity and without loss of 
generality, we assume a complex baseband model.

\begin{subsection}{Dispersion Precompensation \& The Transmission Medium}
\label{subsec:medium}
As explained in Sec.~\ref{sec:introduction}, the proposed scheme is
aimed at short-range fiber-optic communications, \textit{e.g.,} over a
distance $< 10$~km. Therefore, to avoid amplified spontaneous-emission
(ASE) noise, we assume an unamplified optical link. Indeed, we assume
that the only source of noise is the photodiode, as discussed in
Sec.~\ref{subsec:photodiode}. 

We assume that chromatic dispersion is precompensated at the
transmitter using a precoder with transfer function
$H(f)=e^{i2\beta_2L\pi^2f^2}$, where $\beta_2$ is the group-velocity
dispersion parameter and $L$ is the fiber
length~\cite{precom0,precom1,precom2,precom3,precom4}.  Therefore, the
transmitted complex-valued waveform over the fiber is
$u\triangleq\mathcal{F}^{-1}\{X(f)H(f)\}$, where $X(\cdot)$ is the
Fourier transform of $x(\cdot)$, the complex-valued input waveform of
the all-pass filter, and $\mathcal{F}^{-1}$ denotes the inverse
Fourier transform operator.  As the precoder is an all-pass filter,
$x(\cdot)$ and $u(\cdot)$ have equal energy, thus dispersion
precompensation does not result in a precoding loss, which is a
drawback of some partial-response coding techniques, \textit{e.g.},
Tomlinson-Harashima precoding~\cite{th1,th2,th3}. 
   
As the optical-fiber length is relatively short, except for power
loss, we dismiss other transmission impairments, \textit{e.g.},
polarization-mode dispersion, Kerr effect, etc.  As a result, the
received optical waveform is $r(\cdot)=\rho x(\cdot)$, where
$\rho\in(0,1]$ is the loss factor, depending on the fiber length and
the operating wavelength. For simplicity in computations, we assume
that $\rho = 1$, which in optical communications is known as
back-to-back transmission.  It should be mentioned that the BER and
the mutual-information figures, discussed in
Sec.~\ref{sec:numerical_simulation}, depend on the received
optical-signal power. Thus, for a given non-zero transmission length,
\textit{i.e.,} with $\rho<1$, the needed launch power can be computed
easily. Examples with $\rho<1$ are presented in
Sec.~\ref{sec:numerical_simulation}.

\begin{figure}
\centering
\includegraphics[scale=0.6666666]{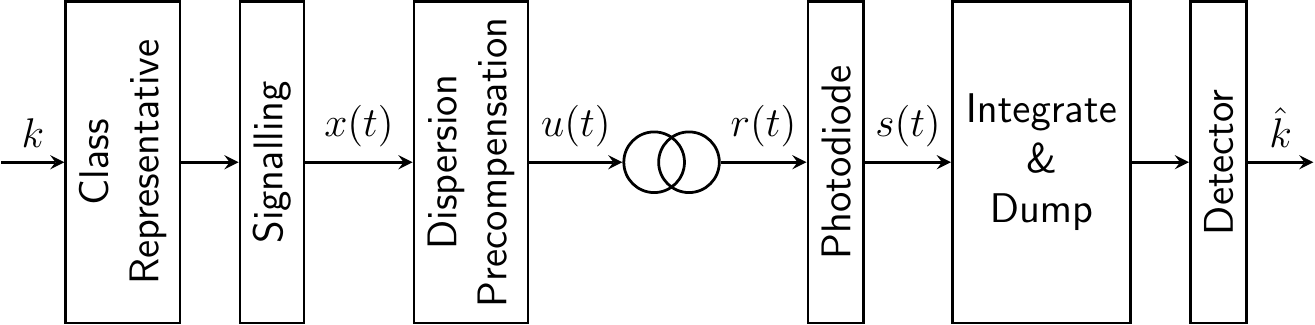}
\caption{The system model; $k\in\{1,\ldots,M\}$, $x(\cdot),u(\cdot),$ 
	and $r(\cdot)$ are
	complex-valued waveforms, and $s(\cdot)$ is a real-valued waveform.
	As the chromatic dispersion is precompensated and the link is 
	unamplified,
	we assume that $r(\cdot)=\rho x(\cdot)$, where
	 $\rho\in(0,1]$.}
\label{fig:system_model}
\end{figure}

\end{subsection}

\begin{subsection}{Tukey Signalling}
\label{subsec:tukey_signalling}

In this section, we describe the ``signalling''
unit of Fig.~\ref{fig:system_model}, which is the central contribution
of this paper. For a positive integer $n$, this unit accepts $n$ 
complex numbers, $x_0,\ldots,x_{n-1}\in\mathbb{C}$,
called the transmitted symbols, or simply the symbols,
and produces the waveform 
$x\in\mathbb{C}^\mathbb{R}$, given as
\begin{equation}
x(t)=\sum_{d=0}^{n-1}x_d w(t-d T),
\label{eq:d2a}
\end{equation} 
where $T\in\mathbb{R}^{>0}$ is the inverse of the baud rate, and $w(\cdot)$
is a real-valued signalling waveform.
Throughout the paper, $n$ denotes the number of transmitted symbols.

In typical
communication systems, $w(\cdot)$ is chosen to be a sinc waveform, 
a raised-cosine waveform, 
a root raised-cosine waveform, etc. 
However, their ``ISI patterns'' 
are complicated, \textit{i.e.,} at a given time $t$ when the ISI is non-zero,
many, if not all, of the transmitted symbols contribute to the
squared-magnitude of $x(t)$. 
As mentioned in Sec.~\ref{sec:introduction},
this increases the detection complexity. To avoid this, we require
that only a few transmitted symbols should interfere at time $t$. 
One way to achieve this is to have
$w(\cdot)$ supported over a relatively short time interval.
However, this reduction in the ``timewidth'' of $w(\cdot)$ increases its 
bandwidth. To tackle this issue, $w(\cdot)$ should have a tapered edge, 
\textit{i.e.,} it should drop slowly and smoothly toward zero. 
A familiar function having this behaviour is the Fourier transform
of the raised-cosine function. 
Note, however, that $w(\cdot)$ must possess this 
tapering property in the time domain, not in the frequency domain.

We propose the use of the following family of waveforms.
For a $\beta\in[0,1]$, let
\begin{align}
w_{\beta}(t)\triangleq
\left\{
\begin{array}{ll}
\frac{2}{\sqrt{4-\beta}}, & \text{if }|t|\leq\frac{(1-\beta)}{2};\\
&\\
\frac{1}{\sqrt{4-\beta}}\left(1-
	\sin\left(\frac{\pi(2|t|-1)}{2\beta}\right)\right),&
		\text{if }\left||t|-\frac{1}{2}\right|\leq\frac{\beta}{2};\\
&\\
0, & \text{otherwise}.
\label{eq:tukey}
\end{array}
\right.
\end{align}
Note that $w_\beta(\cdot)$ has unit energy, \textit{i.e.},
	$\int_{-\infty}^{\infty}w_\beta^2(t)\text{d} t=1$, and is supported
over a time interval of duration $1+\beta$.
This waveform is known in spectrum estimation
as the 
\textit{cosine-tapered window} or the 
\textit{Tukey window}, named
after mathematician and founder of 
``modern spectrum estimation''~\cite{tukey2}, John W.~Tukey,
who suggested them as a combination of 
\textit{rectangular} and \textit{Hann} 
windows~\cite{tukey4,tukey1,tukey3}. 
Indeed, $w_{0}(\cdot)$ and $w_{1}(\cdot)$ are rectangular and Hann windows,
respectively.   
To avoid introducing a new name for 
$w_{\beta}(\cdot)$ as a signalling waveform, we refer to it as a
\textit{Tukey waveform}. 
Fig.~\ref{fig:tukey} shows  
Tukey waveforms for two different $\beta$ values.

\begin{figure}
\centering
\includegraphics[scale=0.6666666]{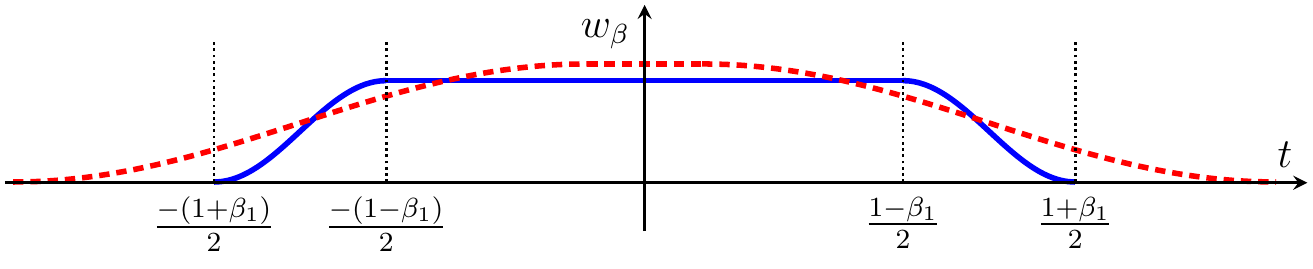}
\caption{Tukey waveforms $w_{\beta_1}(\cdot)$ (solid) and $w_{\beta_2}(\cdot)$ (dashed); 
	$\beta_1<\beta_2$.}
\label{fig:tukey}
\end{figure}

We set $w(\cdot)$ in (\ref{eq:d2a}) to be a dilated
Tukey waveform;
in particular,
$w(t)=w_\beta\left(\frac{t}{T}\right)$.
With this choice of $w(\cdot)$, 
at any time $t$ within the support of $x(\cdot)$, 
either exactly one or exactly two of the transmitted
symbols contribute to $|x(t)|^2$. This property
facilitates the recovery of phase information
from the induced ISI. 
Accordingly, we define two types of time intervals,
to be used in later sections, as follows. 
For $k\in\{0,\ldots,n-1\}$, $x(t)$ depends
only on $x_k$ whenever $|t-kT|\leq\frac{(1-\beta)T}{2}$.
Thus, we define the $k^\text{th}$ ISI-free interval as 
\begin{equation}
\mathcal{Y}_k\triangleq\left[\left(k-\frac{1-\beta}{2}\right)T~,~
	\left(k+\frac{1-\beta}{2}\right)T\right].
\label{eq:interval_y}
\end{equation} 
Similarly, for $\ell\in\{0,\ldots,n-2\}$,
$x(t)$ depends on both $x_\ell$ and $x_{\ell+1}$ whenever 
$\left|t-(\ell+\frac{1}{2})T\right|<\frac{1}{2}\beta T$.
Therefore, we define the $\ell^\text{th}$ ISI-present interval as
\begin{equation}
\mathcal{Z}_\ell\triangleq\left(\left(\ell+\frac{1-\beta}{2}\right)T~,~
	\left(\ell+\frac{1+\beta}{2}\right)T\right).
\label{eq:interval_z}
\end{equation}
The ISI-free and ISI-present intervals are shown (for $n=3$) 
in Fig.~\ref{fig:interval}.

\begin{figure}
\centering
\includegraphics[scale=0.666666]{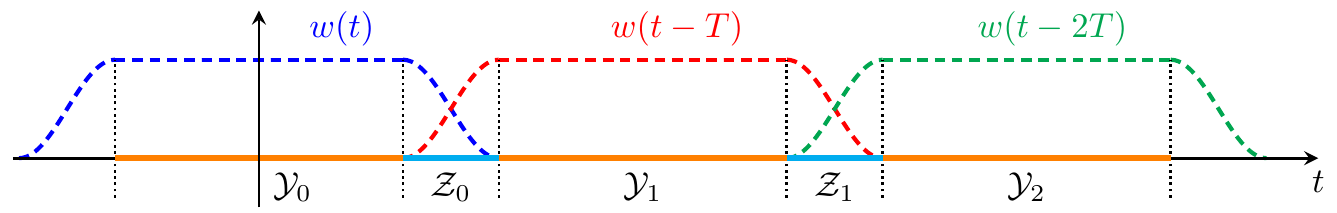}
\caption{ISI-free and ISI-present intervals for $n=3$.}
\label{fig:interval}
\end{figure}

A Tukey waveform $w_\beta(\cdot)$ may be generated by passing a
rectangular pulse
\[
\Pi(t)\triangleq \begin{cases}
		1, &\text{if }|t|\leq\frac{1}{2};\\
			0,&\text{otherwise}
	\end{cases}
\]
through a linear time-invariant (LTI) filter with impulse response
\[
       h_\beta(t)\triangleq \begin{cases}
		\frac{\pi}{\beta\sqrt{4-\beta}}
		\cos\left(\frac{\pi}{\beta}t\right), &\text{if }|t|\leq
			\frac{\beta}{2};\\
			0,&\text{otherwise,}
	\end{cases} \]
as shown in Fig.~\ref{fig:filter}.
In particular, $w_\beta(\cdot)=h_\beta(\cdot)\ast\Pi(\cdot)$, where 
$\ast$ denotes convolution. As will be discussed
in Sec.~\ref{sec:complexity},
this fact can be exploited for waveform
generation at the transmitter.
\begin{figure}
\centering
	\includegraphics[scale=0.6666666]{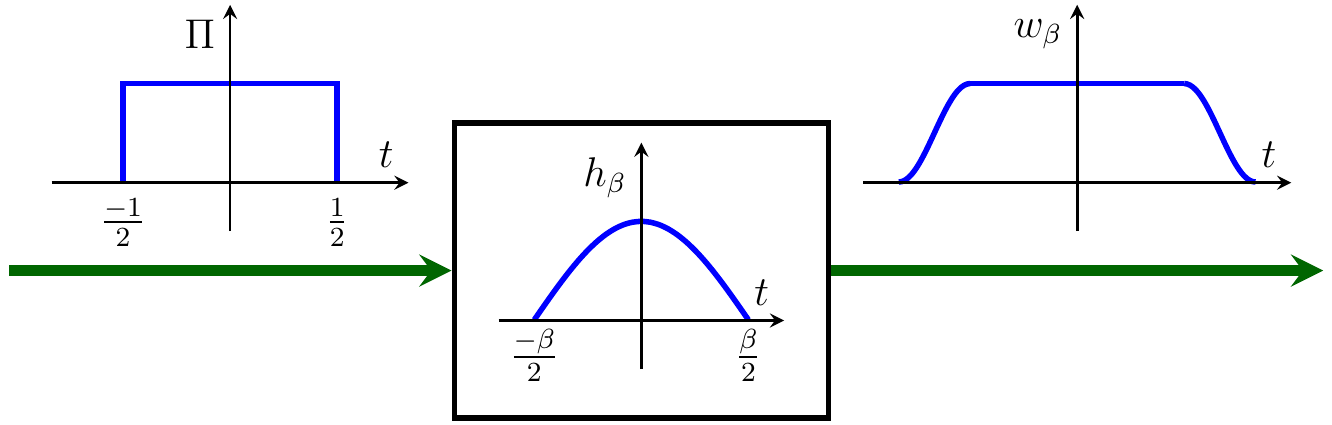}
	\caption{The Tukey waveform $w_\beta(\cdot)$ is the output of an LTI filter
	with impulse response $h_\beta(\cdot)$ to the input waveform $\Pi(\cdot)$.}
	\label{fig:filter}
\end{figure}

As the waveforms $w_\beta(\cdot)$ are strictly time-limited, they
cannot be bandlimited.
The Fourier transform of $w_\beta(\cdot)$, denoted as $W_\beta(\cdot)$, is
\begin{equation*}
W_\beta(f)\triangleq
\left\{
\begin{array}{ll}
\frac{\pi}{2\sqrt{4-\beta}}\text{sinc}\left(\frac{1}{2\beta}\right),&
\text{if }f=\frac{\pm1}{2\beta};\\
&\\
\frac{2}{\sqrt{4-\beta}}\text{sinc}(f)\frac{\cos(\pi\beta f)}{1-(2\beta f)^2},&
\text{otherwise}.
\end{array}
\right.
\end{equation*}
Fig.~\ref{fig:Fourier} shows $W_\beta(\cdot)$ for two different $\beta$ values.
Tables~\ref{tab:95_bandwidth} and~\ref{tab:90_bandwidth} give the 
bandwidth of $w_\beta(\cdot)$ which contains, 
respectively, $95\%$ and $90\%$ of the 
total energy. For these energy percentiles, the out-of-band signal energy 
is $12.79~$dB and $9.54~$dB less than the in-band energy, respectively. 
One observes that, although not strictly bandlimited,
the bandwidth of $w_{\beta}(\cdot)$ 
is close to $1/2$ for large values of $\beta$.
For example, $95\%$ of the $w_{0.9}(\cdot)$
energy is contained within a spectral band of 
length $0.575$, \textit{i.e.}, a $15\%$ overhead compared to
the minimum bandwidth required for Nyquist signalling.
Indeed, the numerical simulations
in Sec.~\ref{sec:numerical_simulation}
support the use of large $\beta$ values in terms of
BER and mutual information.

Note that $w_\beta(\cdot)$ is not orthogonal to its unit-shift
replica, \textit{i.e.},
$\int_{-\infty}^{\infty}w_\beta(t)w_\beta(t-1)\text{d} t\neq 0$. This lack
of orthogonality must be taken into account when computing the average
power of the waveform $x(\cdot)$.  Luckily, in the case when the
symbols are independent and identically distributed (i.i.d.),
Theorem~\ref{theorem} shows that this power is indeed given as the
mean squared value of the symbol magnitudes.
\begin{theorem}
\label{theorem}
For a positive integer $m$ and $P \in \mathbb{R}^{>0}$,
let $\lambda_0,\ldots,\lambda_{m-1}$ be  
i.i.d. zero-mean complex random variables such that  
$\expect{|\lambda_0|^2}=P$ and $\var{|\lambda_0|^2}<\infty$.
Furthermore, let 
\begin{equation*}
\Lambda_m(t)=\sum_{j=0}^{m-1}\lambda_jw_\beta\left(\frac{t}{T}-j\right).
\end{equation*}
Then,
\begin{equation*}
	\frac{1}{mT}\int_{-\infty}^{\infty}|\Lambda_m(t)|^2\text{d} t
	\overset{p}{\rightarrow}P
\quad\text{as }m\rightarrow\infty,
\end{equation*}
\textit{i.e.}, the power of $\Lambda_m(\cdot)$ converges to $P$ in probability.
\end{theorem}
\begin{IEEEproof}
	See Appendix.
\end{IEEEproof}
We remark that finiteness of $\var{|\lambda_0|^2}$ is a very mild
condition which is true for all finite signal constellations and
indeed most of the usual distributions over an infinite range.
\begin{table}
\centering
\caption{The bandwidth containing $95\%$ of the $w_{\beta}(\cdot)$ 
energy and its
overhead compared to 
the minimum bandwidth for Nyquist signalling}
\label{tab:95_bandwidth}
\begin{tabular}{|c|c|c|c|c|c|}
\hline
$\beta$ & bandwidth & overhead & $\beta$ & bandwidth & overhead\\
\hline
\hline 
$0.1$ & $1.477$ & $195.4\%$ & $0.3$ & $0.788$ & $57.6\%$\\
\hline
$0.5$ & $0.668$ & $33.6\%$ & $0.7$ & $0.613$ & $22.6\%$\\
\hline
$0.8$ & $0.592$ & $18.4\%$ & $0.9$ & $0.575$ & $15\%$ \\
\hline
\end{tabular}
\end{table}

\begin{table}
\centering
\caption{The bandwidth containing $90\%$ of the $w_{\beta}(\cdot)$ 
energy and its
overhead compared to the minimum bandwidth for Nyquist signalling}
\label{tab:90_bandwidth}
\begin{tabular}{|c|c|c|c|c|c|}
\hline
$\beta$ & bandwidth & overhead & $\beta$ & bandwidth & overhead\\
\hline
\hline 
$0.1$ & $0.706$ & $41.2\%$ & $0.3$ & $0.612$ & $22.4\%$\\
\hline
$0.5$ & $0.56$ & $12\%$ & $0.7$ & $0.522$ & $4.4\%$\\
\hline
$0.8$ & $0.505$ & $1\%$ & $0.9$ & $0.49$ & $-2\%$ \\
\hline
\end{tabular}
\end{table}

\begin{figure}
\centering
\includegraphics[scale=0.6666666]{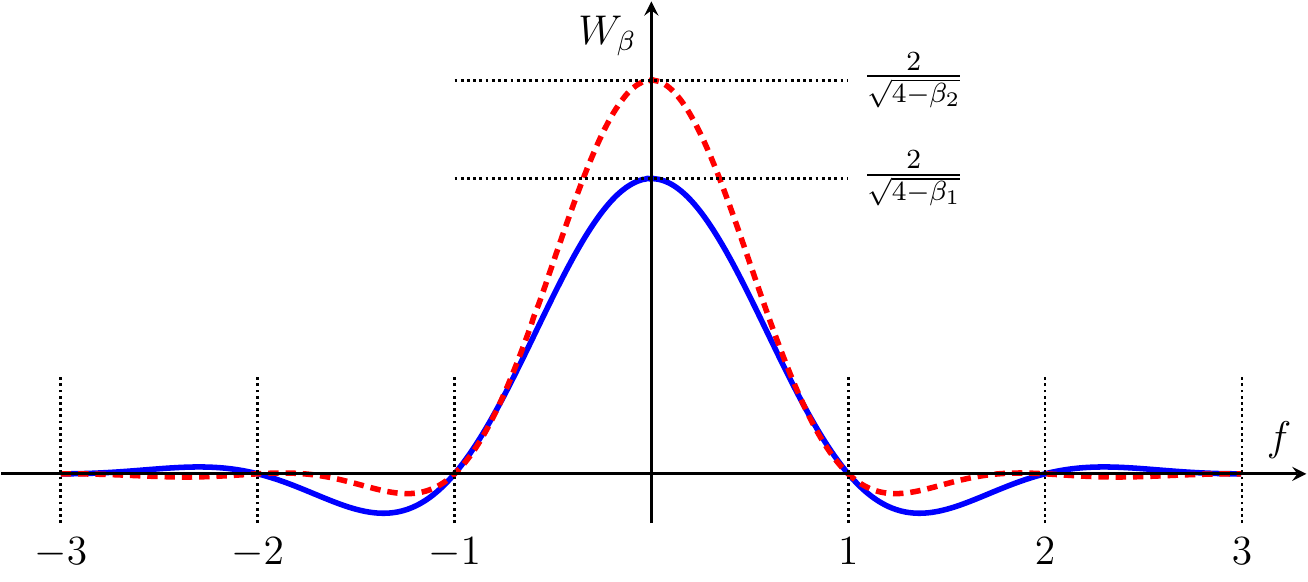}
\caption{Fourier transforms of $w_{\beta_1}(\cdot)$
	(solid) and $w_{\beta_2}(\cdot)$ (dashed); $\beta_1<\beta_2$.}
\label{fig:Fourier}
\end{figure}
\end{subsection}

\begin{subsection}{Photodiode, the Only Source of Noise}
\label{subsec:photodiode}

The received optical waveform, $r(\cdot)$,
is converted to an electrical signal by a photodiode.
In our model, the photodiode is the only source of noise.
The photodiode has a gain which, for simplicity, is assumed to be unity
in this section. However, the actual gain is taken into account
in the numerical simulations presented in Sec.~\ref{sec:numerical_simulation}. 
The output of the photodiode is a real-valued waveform $s(\cdot)$, such that
\begin{equation}
s(t)=|r(t)|^2+|r(t)|\shot{t}+\thermal{t},
\label{eq:s}
\end{equation}
where $\shot{\cdot}$ and $\thermal{\cdot}$ are independent zero-mean
white Gaussian random processes with constant two-sided
power spectral densities (PSDs) 
$\sshot^2$ and $\sthermal^2$, respectively. The $|r(t)|\shot{t}$ and 
$\thermal{t}$ terms are known, respectively, as shot noise and 
thermal noise~\cite{agrawal}. It should be mentioned that photodiodes have
additional practical deficiencies, such as, \textit{e.g.,} dark current.
However, we assume that such effects contribute negligibly compared
to the noise terms in (\ref{eq:s}).  
\end{subsection}

\begin{subsection}{Integrate \& Dump}
\label{subsec:integrate_and_dump}

In this section, we discuss the integrate-and-dump unit
in Fig~\ref{fig:system_model}. 
As noted in Sec.~\ref{subsec:tukey_signalling},
at any time $t$ within the support of $x(\cdot)$,
either exactly one or exactly two of the transmitted symbols 
contribute to $|x(t)|^2$.
The integrate-and-dump unit integrates its input waveform over 
each $\mathcal{Y}_k$ and $\mathcal{Z}_\ell$ interval, producing $y_k$ and
$z_\ell$, respectively, where $k\in\{0,\ldots,n-1\}$ and 
$\ell\in\{0,\ldots,n-2\}$.
More precisely,
\begin{equation}
	y_k\triangleq\int_{\mathcal{Y}_k}s(t)\text{d} t,
\label{eq:y}
\end{equation}
and
\begin{equation}
	z_\ell\triangleq\int_{\mathcal{Z}_\ell}s(t)\text{d} t.
\label{eq:z}
\end{equation}
We expand (\ref{eq:y}) as follows.
Let $\alpha\triangleq\frac{2}{\sqrt{4-\beta}}$; then, 
by using $r(\cdot)=x(\cdot)$ (see Sec.~\ref{subsec:medium}),
we get
\begin{equation}
y_k= \alpha^2(1-\beta)T|x_k|^2+\alpha|x_k|n_k+m_k,
\label{eq:simplified_y}
\end{equation}
where
\begin{equation*}
	n_k\triangleq\int_{\mathcal{Y}_k}\shot{t}\text{d} t\sim
	\mathcal{N}(0,\sshot^2(1-\beta)T),
\end{equation*}
and
\begin{equation*}
	m_k\triangleq\int_{\mathcal{Y}_k}\thermal{t}\text{d} t\sim
	\mathcal{N}(0,\sthermal^2(1-\beta)T).
\end{equation*}
Note that $\expect{n_kn_{k'}}=\expect{m_km_{k'}}=0$, for any $k'\neq k$, where
$k'\in\{0,\ldots,n-1\}$.

We expand (\ref{eq:z}) as follows.  Let
$\psi:\mathbb{C}^2\rightarrow\mathbb{R}$ be defined as
\begin{equation*}
\psi(v,w)=\frac{1}{4}|v+w|^2+\frac{1}{8}|v-w|^2,
\end{equation*}
for any $v$ and $w\in\mathbb{C}$.
Then, (\ref{eq:z}) can be simplified as
\begin{equation}
z_\ell=\alpha^2\beta T \psi(x_\ell,x_{\ell+1})+
	\alpha\sqrt{\psi(x_\ell,x_{\ell+1})}p_\ell+q_\ell,
\label{eq:simplified_z}
\end{equation}
where
$p_\ell\sim\mathcal{N}(0,\beta T \sshot^2)$,
and
$q_\ell\sim\mathcal{N}(0,\beta T \sthermal^2)$.
Similarly, $\expect{p_\ell p_{\ell'}}=\expect{q_\ell q_{\ell'}}=0$, for any
$\ell'\neq\ell$.
Note that for any $k$ and $k'\in\{0,\ldots,n-1\}$ and
for any $\ell$ and $\ell'\in\{0,\ldots,n-2\}$, the
four random variables $n_k, m_{k'}, p_\ell,$ and $q_{\ell'}$ are
mutually independent.

\end{subsection}

\begin{subsection}{Equivalence Classes}
\label{subsec:equivalence_classes}

In this section, we describe the first transmitter
block in Fig.~\ref{fig:system_model},
\textit{i.e.}, choice of class representative.
Let $\Upsilon:\mathbb{C}^n\rightarrow\mathbb{R}^{n}\times\mathbb{R}^{n-1}$
denote the function that maps a vector
$\bm{x}=(x_0,\ldots,x_{n-1})\in\mathbb{C}^n$
at the input of the signalling block to the corresponding output of
the integrate-and-dump block in the absence of noise and loss
(i.e.,  in a back-to-back configuration with $L=0$),
as shown in Fig.~\ref{fig:g_function}. Specifically, if 
$x(t)=\sum_{d=0}^{n-1}x_dw(t-dT)$ then 
$\Upsilon(\bm{x})=(\bm{y},\bm{z})$,
where $\bm{y}\in\mathbb{R}^{n}$ is such that
\begin{equation*}
	\bm{y}[k]=\int_{\mathcal{Y}_k}|x(t)|^2 \text{d} t,
\end{equation*}
for $k\in\{0,\ldots,n-1\}$, and $\bm{z}\in\mathbb{R}^{n-1}$
is such that
\begin{equation*}
	\bm{z}[\ell]=\int_{\mathcal{Z}_\ell}|x(t)|^2 \text{d} t,
\end{equation*}
for $\ell\in\{0,\ldots,n-2\}$.

We define an equivalence relation on 
$\mathbb{C}^n$ as follows. Two vectors $\bm{x}$ and 
$\bm{\tilde{x}}\in\mathbb{C}^n$ are said to be
\textit{square-law identical}, denoted
$\bm{x}\relates\bm{\tilde{x}}$,
if and only if $\Upsilon(\bm{x})=\Upsilon(\bm{\tilde{x}})$. 
If $\bm{x}$ and $\bm{\tilde{x}}$ are not 
square-law identical, they are said to be 
\textit{square-law distinct}, denoted
$\bm{x}\nrelates\bm{\tilde{x}}$.
Since the relation $\relates$ is indeed an equivalence relation,
it partitions $\mathbb{C}^n$ into disjoint equivalence classes. 

\begin{example}
Let $\bm{x}=(1,i,1,-1)$ and $\bm{\tilde{x}}=(1,i,-1,1)$.
For both of these vectors we have 
$\Upsilon(\bm{x})=\Upsilon(\bm{\tilde{x}})=(\bm{y},\bm{z})$ where
$\bm{y}=\alpha^2(1-\beta)T(1,1,1,1)$
and $\bm{z}=\alpha^2\beta T(\frac{3}{4},\frac{3}{4},\frac{1}{2})$. Therefore, 
$\bm{x}\relates\bm{\tilde{x}}$, \textit{i.e.}, $\bm{x}$ and $\bm{\tilde{x}}$
are square-law identical. However, for $\bm{\hat{x}}=(1,-1,1,i)$ we have 
$\Upsilon(\bm{\hat{x}})=(\bm{\hat{y}},\bm{\hat{z}})$ where 
$\bm{\hat{y}}=\bm{y}$, but $\bm{\hat{z}}=\alpha^2\beta T(\frac{1}{2},
\frac{1}{2},\frac{3}{4})\neq\bm{z}$; therefore, $\bm{x}\nrelates\bm{\hat{x}}$. 
\end{example}

For any positive integer $M$,
let $\mathcal{S}$ be a set of cardinality $M$,
whose elements are complex-valued vectors
of length $n$, called \textit{symbol blocks}, which are  
square-law distinct.
In other words, 
$\mathcal{S}=\left\{\bm{x}_1,\ldots,\bm{x}_M\right\}\subset
	\mathbb{C}^n$ such that 
$\bm{x}_k\nrelates\bm{x}_\ell$ if $k\neq\ell$.
Then, the class-representative unit outputs
the symbol block $\bm{x}_k$ if its input is $k\in\{1,\ldots,M\}$. 
Note that the entries of symbol blocks are in fact the transmitted
symbols, i.e., the set $\mathcal{S}$ forms a
signal constellation in $n$ complex dimensions ($2n$ real dimensions).
Furthermore, note that the spectral efficiency of this 
communication scheme cannot exceed $\frac{1}{n}\log_2(M)$ bit/sec/Hz.

\begin{figure}
\centering
\includegraphics[scale=0.666666]{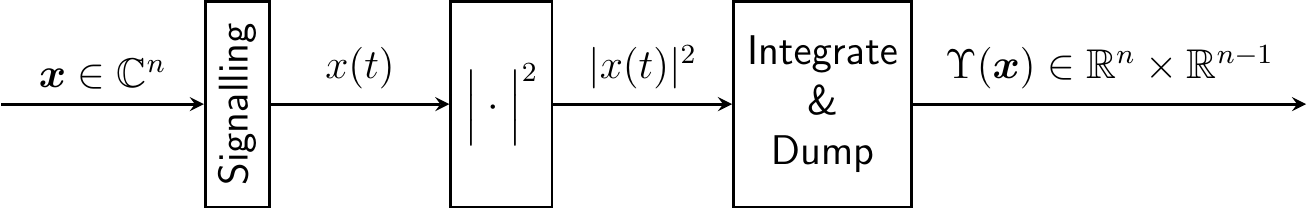}
\caption{The $\Upsilon$ function maps the input of the signalling block to 
the corresponding output of the integrate-and-dump block in the absence
of noise. The noisy photodiode of Fig.~\ref{fig:system_model}
has been replaced with an ideal $|\cdot|^2$ block.}
\label{fig:g_function}
\end{figure}
\end{subsection}

\begin{subsection}{Maximum-Likelihood Block Detection}
\label{sec:ml}

The last block of Fig.~\ref{fig:system_model} is the detector.
In practice, the choice of
detection rule and its algorithm will depend on hardware limitations
and performance requirements. Through out this paper, 
maximum-likelihood (ML) block detection
is chosen as the detection rule. In other words,
if $\bm{y}=(y_0,\ldots,y_{n-1})$ and $\bm{z}=(z_0,\ldots,z_{n-2})$
are the buffered outputs of the integrate-and-dump unit,  
the detector chooses $\hat{k}\in\{1,\ldots,M\}$ as the transmitted 
index if and only if 
\begin{equation*}
\hat{k}=\underset{d\in\{1,\ldots,M\}}{\arg\max}
	f\left(\bm{y},\bm{z}\mid\bm{x}_d\right). 
\end{equation*}
For any $k\in\{0,\ldots,n-1\}$, 
(\ref{eq:simplified_y}) implies that, given $x_k$,
\begin{equation}
y_k\sim
\mathcal{N}\Big(
	\alpha^2(1-\beta)T|x_k|^2,
	(1-\beta)T\left(\alpha^2|x_k|^2\sshot^2+\sthermal^2\right)
			\Big).
\label{eq:y_likelihood}
\end{equation} 
Furthermore, for any $\ell\in\{0,\ldots,n-2\}$, (\ref{eq:simplified_z}) 
implies that,
given $x_\ell$ and $x_{\ell+1}$,
\begin{equation}
z_\ell\sim
\mathcal{N}\Big(
\alpha^2\beta T\psi(x_\ell,x_{\ell+1}),
	\beta T\left(\alpha^2\psi(x_\ell,x_{\ell+1})\sshot^2+\sthermal^2\right)
	\Big).
\label{eq:z_likelihood}
\end{equation} 
Thus,
\begin{align}
f(\bm{y},\bm{z}\mid\bm{x}_d)&=f(\bm{y}\mid\bm{x}_d)
				f(\bm{z}\mid\bm{x}_d)\nonumber\\
&=\prod_{k=0}^{n-1}f(y_k\mid \bm{x}_d[k])
	\prod_{\ell=0}^{n-2}f(z_\ell\mid \bm{x}_d[\ell],\bm{x}_d[\ell+1]),
\label{eq:factoring}
\end{align}  
where $f(y_k\mid \bm{x}_d[k])$ and
$f(z_\ell\mid \bm{x}_d[\ell],\bm{x}_d[\ell+1])$ can be
computed from (\ref{eq:y_likelihood}) 
and (\ref{eq:z_likelihood}), respectively.
Fig.~\ref{fig:factor_graph} is the factor-graph 
representation~\cite{factor_graph} of (\ref{eq:factoring}). 
Note that the variances of $y_k$ and $z_\ell$, given in
(\ref{eq:y_likelihood})
and (\ref{eq:z_likelihood}), are 
functions of the transmitted symbols; therefore, 
minimum Euclidean-distance
detection is \emph{not} equivalent to ML block detection.
\begin{figure}
\centering
\includegraphics[scale=0.6666666]{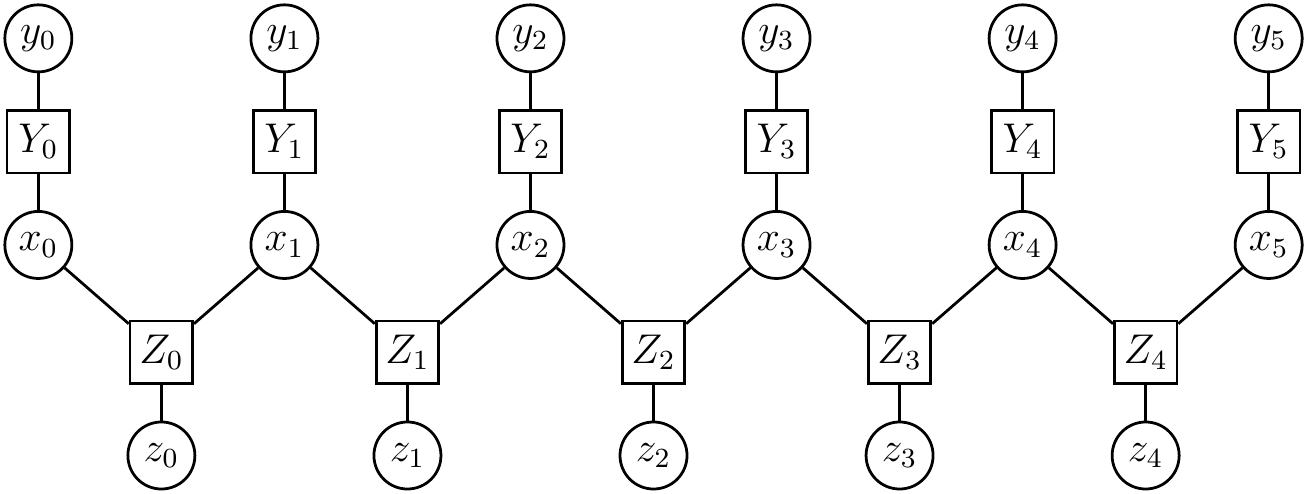}
\caption{The factor graph of the ML block detection for $n=6$.
	For $k\in\{0,\ldots,5\}$, the $Y_k$ factor node represents
	$f(y_k\mid x_k)$. Similarly, for $\ell\in\{0,\ldots,4\}$,
	the $Z_\ell$ factor node represents $f(z_\ell\mid x_\ell,x_{\ell+1})$.}
\label{fig:factor_graph}
\end{figure}
\end{subsection}
\end{section}

\begin{section}{Numerical Simulation}
\label{sec:numerical_simulation}

In this section, the communication scheme proposed in 
Sec.~\ref{sec:system_model} is verified by numerical simulations.

\begin{subsection}{The Photodiode}
\label{subsec:the_photodiode}

We assume the use of an InGaAs avalanche photodiode (APD),
as among different types of APDs, these have high bandwidth.
In this case, the two-sided PSD of the thermal noise, $\thermal{\cdot}$, 
is~\cite{agrawal}
\begin{equation*}
\sthermal^2=\frac{2\mathsf{k}T_k}{R_L},
\end{equation*}
where $\mathsf{k}$ is the Boltzmann constant, $T_k$ is the temperature, and 
$R_L$ is the external load resistance. 
Furthermore, for the received optical complex-valued waveform $r(\cdot)$,
the shot noise is $|r(t)|\shot{t}$, where
the two-sided PSD of $\shot{\cdot}$ is
\begin{equation*}
\sshot^2=eM_\text{APD}^2FR_\text{APD},
\end{equation*} 
where $e$ is the unit charge, $M_\text{APD}$ is the APD gain, $F$ is the excess
noise factor, and $R_\text{APD}$ is the responsivity of APD. 
Table~\ref{tab:values} gives the values for these parameters used in 
simulations.  
\begin{table}
\centering
\caption{Parameter values used in the numerical simulation}
\label{tab:values}
\begin{tabular}{|c|c|c|}
\hline
\textbf{Parameter} & \textbf{Value} &
$\begin{array}{c}\text{\textbf{Typical Range}}\\ 
\text{\textbf{(InGaAs APD)}}\end{array}$\\
\hline
\hline
Temperature $(T_k)$ & $300~$K & \\
\hline
Load Resistance $(R_L)$ & $15~\Omega$ &\\
\hline
APD Gain $(M_\text{APD})$ & $20$ & $10-40$\\
\hline 
$\begin{array}{c}
\text{Enhanced Responsivity}\\
(M_\text{APD}R_\text{APD})
\end{array}$ & $10~$mA/mW & $5-20$\\
\hline
$k-$factor $(k_\text{APD})$ & $0.6$ & $0.5-0.7$\\
\hline
Excess Noise Factor $(F)$ & $12.78$ &
$\begin{array}{c}
\text{a function of $M_\text{APD}$}\\
\text{and $k_\text{APD}$}
\end{array}$\\
\hline 
\end{tabular}
\end{table}

\end{subsection}

\begin{subsection}{The Transmitted Symbols}
\label{subsec:transmitted_symbols}

The transmitted symbols are chosen from a finite constellation,
$\mathcal{K}$, five of which are shown in Fig.~\ref{fig:constellations}.
Indeed, $\mathcal{S}\subset\mathcal{K}^n$, 
and since the symbol blocks must be square-law distinct,
there is a loss 
in the maximum achievable rate in the proposed scheme, compared to 
its coherent-detection counterpart. Table~\ref{tab:classes} provides
the number of 
equivalent classes of each possible size for  
different values of $n$, and for $2$-ring/$4$-ary phase constellation
(see Fig.~\ref{fig:24}). 
For example,  
for $n=4$ there are $432$ equivalence classes, among  which
$192$ have size $8$. As there are 
$|\mathcal{K}|^n=8^4$ possible vectors of length $4$ over $\mathcal{K}$,
the lost rate is 
$\frac{1}{4}\log_2\left(\frac{8^4}{432}\right)=0.94$ bits
per symbol.
Table~\ref{tab:classes} shows that
the larger is $n$, the smaller
is the rate loss. However, this reduction in rate loss comes with
a higher detection complexity. Specifically, the number of equivalence classes
grows exponentially with $n$, which itself increases the complexity of
the ML block decoder. 

It should be noted that a relatively small
$n$ suffices to achieve a practical target BER. This is because
the transmission scheme is usually the inner-most layer 
of a larger communication system,
which typically also employs an outer error-correcting code.  
\newsavebox{\tempbox}
\newsavebox{\tempboxBig}
\begin{figure}
\centering
\sbox{\tempbox}{\includegraphics[scale=0.465]{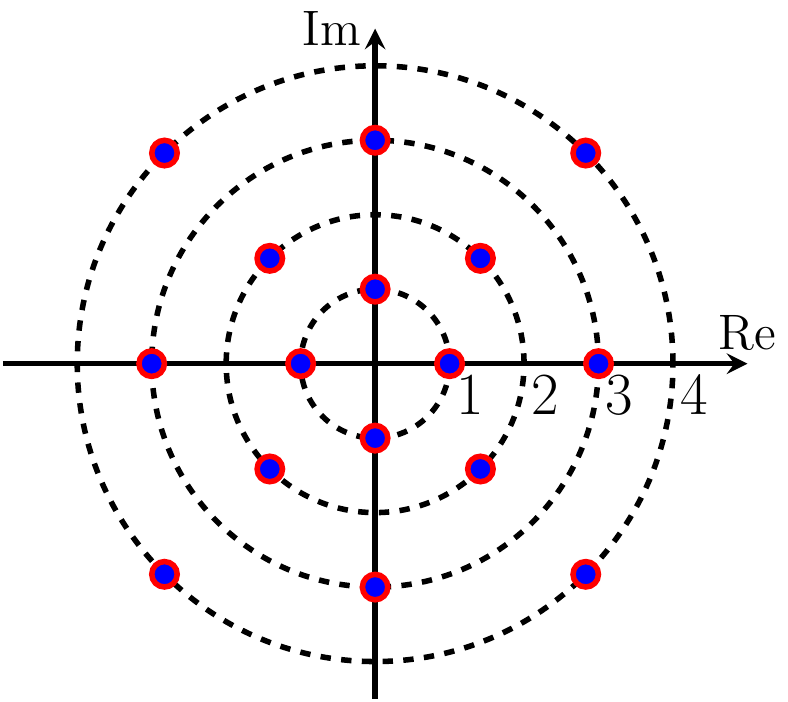}}
\hspace*{-3.5cm}
\subfloat[\label{fig:4psk}]{\vbox to \ht\tempbox{%
\vfil
\includegraphics[scale=0.465]{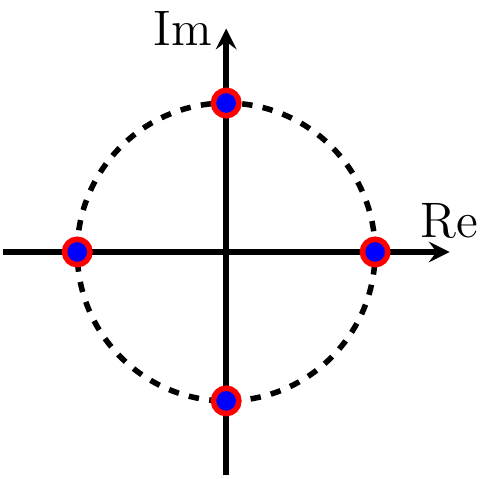}
\vfil}}\hspace*{-6.5cm}
\subfloat[\label{fig:24}]{\vbox to \ht\tempbox{%
\vfil
\includegraphics[scale=0.465]{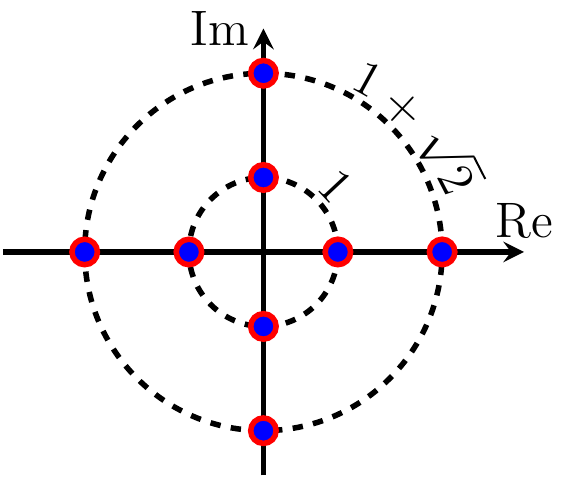}
\vfil}}\hspace*{-3cm}
\subfloat[\label{fig:44}]{\usebox{\tempbox}}\hspace*{-3.2cm}\\
\sbox{\tempboxBig}{\includegraphics[scale=0.29]{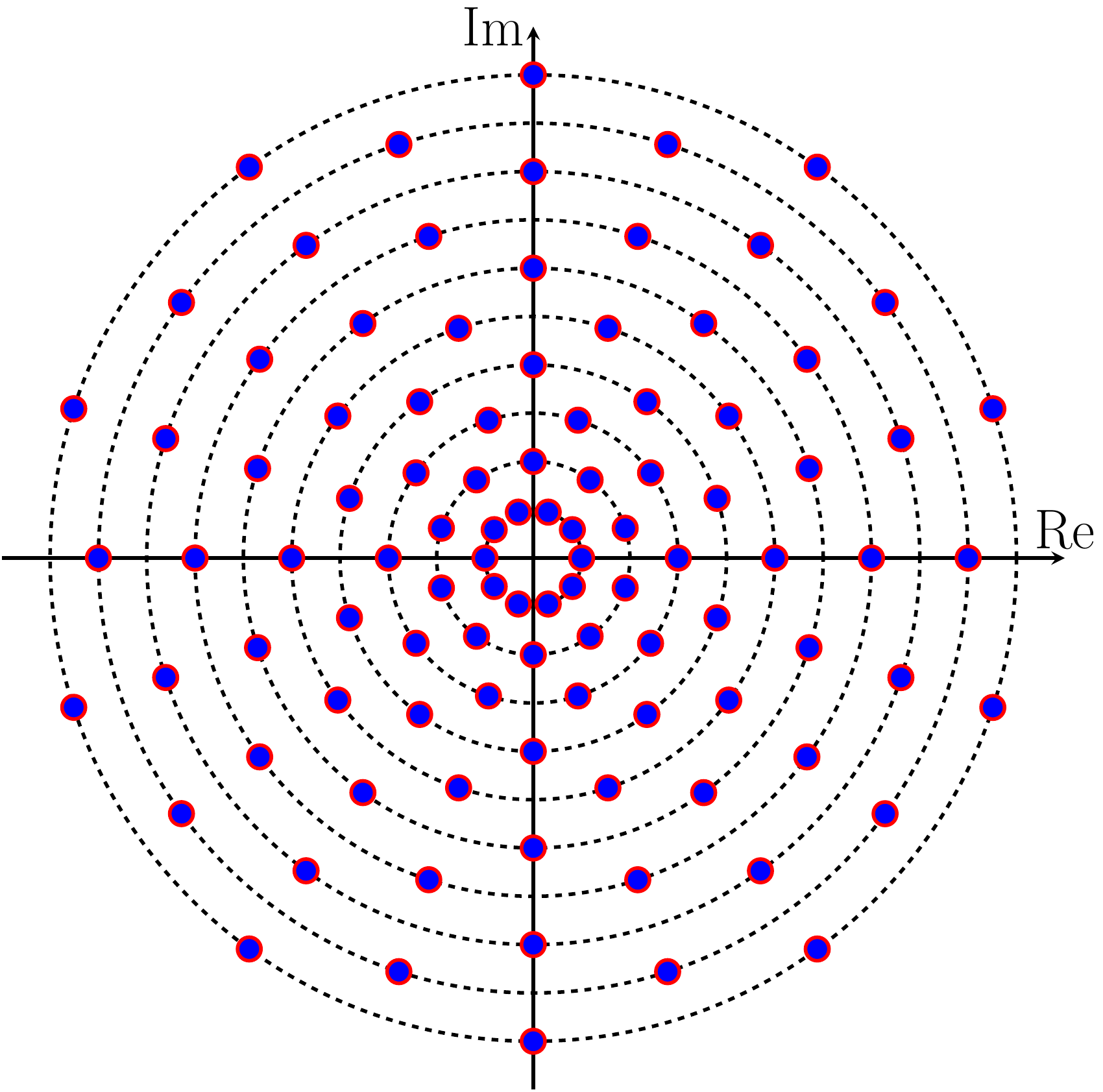}}
\hspace*{-6cm}
\subfloat[\label{fig:88}]{\vbox to \ht\tempboxBig{%
\vfil
\includegraphics[scale=0.29]{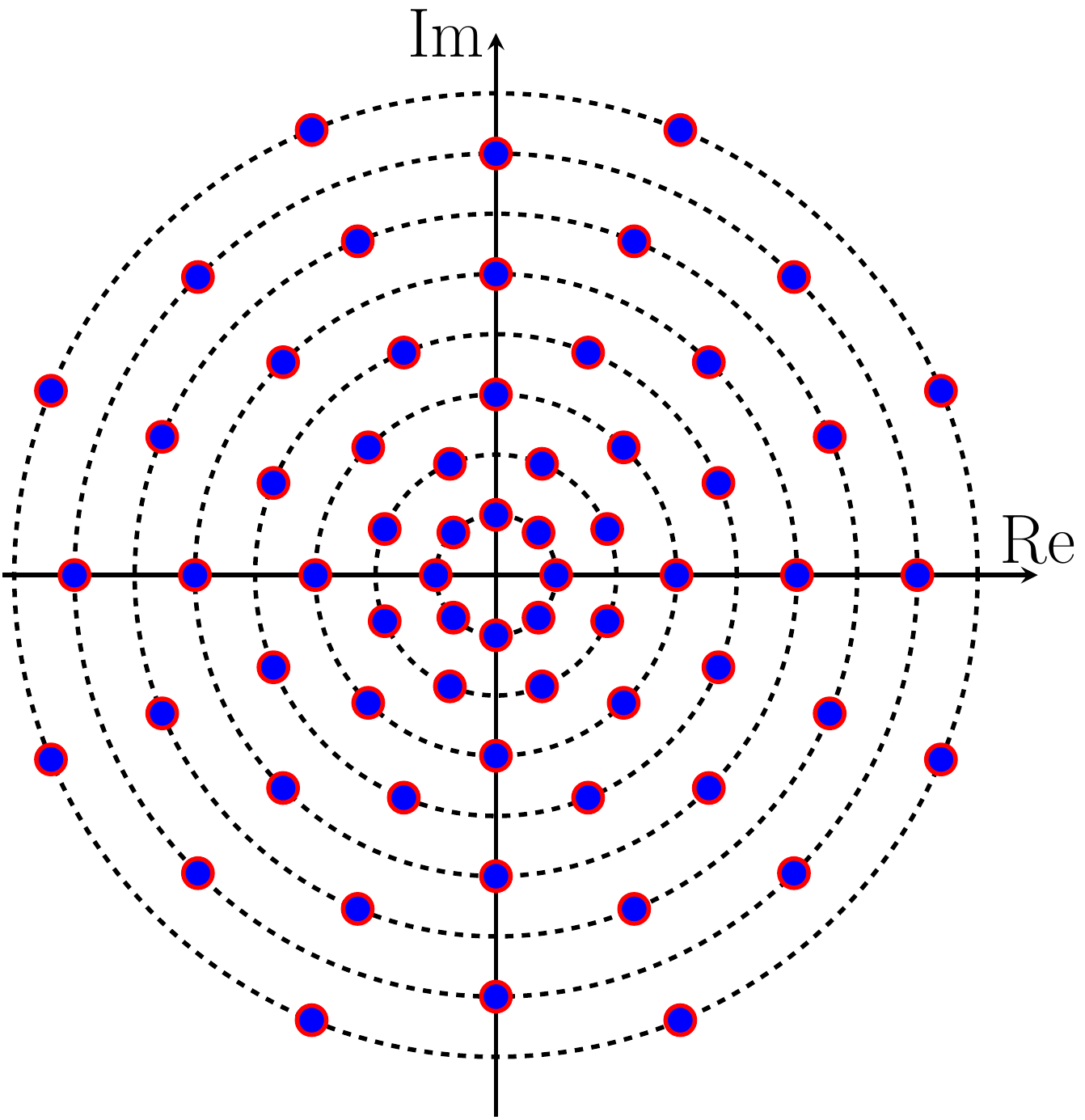}
\vfil}}\hspace*{-2.5cm}
\subfloat[\label{fig:1010}]{\usebox{\tempboxBig}}\hspace*{-3.4cm}
\caption{Five choices for $\mathcal{K}$, the constellation from which the 
	transmitted symbols come: (a) 4-PSK (phase shift keying),
	(b) 2-ring/4-ary phase, 
	(c) 4-ring/4-ary phase,
	(d) 8-ring/8-ary phase,
	(e) 10-ring/10-ary phase constellations. The ratio of the outer-ring
	radius to the inner one in (b) is $1+\sqrt{2}$.
	In (c)--(e) the rings are equi-spaced.}
\label{fig:constellations}
\end{figure}

\begin{table}
\centering
\caption{Number of classes for each class size for $2$-ring/$4$-ary phase
	constellation}
\label{tab:classes}

\begin{tikzpicture}[box/.style={minimum width=9ex,minimum height=1.5ex,inner sep=1pt,align=right}]
\node[inner sep=0pt,outer sep=0pt,minimum width=20ex,minimum height=4ex] (A) {};
\node[anchor=north east,inner sep=2pt] at (A.north east) {$n$};
\node[anchor=south west,inner sep=2pt] at (A.south west) {class size};
\draw (A.north west) -- (A.south east);
\node[box,anchor=west,minimum height=4ex] (B) at (A.east) {3};
\node[box,anchor=west,minimum height=4ex] (C) at (B.east) {4};
\node[box,anchor=west,minimum height=4ex] (D) at (C.east) {5};
\node[box,anchor=west,minimum height=4ex] (E) at (D.east) {6};
\node[box,anchor=west,minimum height=4ex] (F) at (E.east) {7};
\node[anchor=north east] (AA) at (A.south east) {\begin{tabular}{r}
4 \\ 8 \\ 16 \\ 32 \\ 64 \\ 128 \\ 256 \end{tabular}};
\node[anchor=north east] (BB) at (B.south east) {\begin{tabular}{r}
32 \\ 32 \\ 8 \end{tabular}};
\node[anchor=north east] (CC) at (C.south east) {\begin{tabular}{r}
128 \\ 192 \\ 96 \\ 16  \end{tabular}};
\node[anchor=north east] (DD) at (D.south east) {\begin{tabular}{r}
512 \\ 1024 \\ 768 \\ 256 \\ 32  \end{tabular}};
\node[anchor=north east] (EE) at (E.south east) {\begin{tabular}{r}
2048 \\ 5120 \\ 5120 \\ 2560 \\ 640 \\ 64  \end{tabular}};
\node[anchor=north east] (FF) at (F.south east) {\begin{tabular}{r}
8192 \\ 24576 \\ 30720 \\ 20480 \\ 7680 \\ 1536 \\ 128 \end{tabular}};
\node[anchor=north east] (AAA) at (AA.south east) {\begin{tabular}{r}Total\end{tabular}};
\node[box,anchor=east] (BBB) at (AAA -| BB.east) {\begin{tabular}{r}72\end{tabular}};
\node[box,anchor=east] (CCC) at (AAA -| CC.east) {\begin{tabular}{r}432\end{tabular}};
\node[box,anchor=east] (DDD) at (AAA -| DD.east) {\begin{tabular}{r}2592\end{tabular}};
\node[box,anchor=east] (EEE) at (AAA -| EE.east) {\begin{tabular}{r}15552\end{tabular}};
\node[box,anchor=east] (FFF) at (AAA -| FF.east) {\begin{tabular}{r}93312\end{tabular}};
\node[anchor=north,align=center] (AAAA) at (A |- AAA.south) {Rate loss (bit/sym)\\$\frac{1}{n} \log_2 \left( \frac{8^n}{\text{Total}} \right)$};
\node[box,anchor=east] (BBBB) at (AAAA -| BB.east) {\begin{tabular}{r}0.94\end{tabular}};
\node[box,anchor=east] (CCCC) at (AAAA -| CC.east) {\begin{tabular}{r}0.81\end{tabular}};
\node[box,anchor=east] (DDDD) at (AAAA -| DD.east) {\begin{tabular}{r}0.73\end{tabular}};
\node[box,anchor=east] (EEEE) at (AAAA -| EE.east) {\begin{tabular}{r}0.68\end{tabular}};
\node[box,anchor=east] (FFFF) at (AAAA -| FF.east) {\begin{tabular}{r}0.64\end{tabular}};
\coordinate (TL) at (A.north west);
\coordinate (BR) at (F.east |- AAAA.south);
\draw (TL) -- (TL -| BR) -- (BR) -- (BR -| TL) -- cycle;
\draw (A.north east) -- (A.north east |- BR);
\draw (A.south west) -- (A.south west -| BR);
\draw (AAA.south -| TL) -- (AAA.south -| BR);
\draw (AAA.north -| TL) -- (AAA.north -| BR);
\draw (B.north east) -- (B.north east |- BR);
\draw (C.north east) -- (C.north east |- BR);
\draw (D.north east) -- (D.north east |- BR);
\draw (E.north east) -- (E.north east |- BR);
\end{tikzpicture}

\end{table}

\end{subsection}
\begin{subsection}{Mutual Information}
\label{subsec:mutual_information}

Fig.~\ref{fig:rates} shows the achievable rate per symbol in different
scenarios, plotted against the received optical power (ROP), computed
by a Monte Carlo method.  To compute the spectral efficiency in
bit/sec/Hz one must choose the measure of bandwidth,
\textit{e.g.}, $90\%$ in-band energy, $95\%$ in-band energy, etc.,
that suits the application restrictions. The spectral efficiency
can then be computed readily. 

In Figs.~\subreference{fig:rates}{fig:24_3}--\subreference{fig:rates}{fig:all},
the transmitted symbol block is chosen independently and uniformly among all
symbol blocks, \textit{i.e.}, class representatives.
In Fig.~\subreference{fig:rates}{fig:uniform}, instead of
square-law distinct symbol blocks, all possible vectors of 
$\mathcal{K}^n$ have been chosen with uniform distribution.
As equivalence classes have different cardinalities 
(see Table~\ref{tab:classes}), 
a uniform distribution over
$\mathcal{K}^n$ induces a non-uniform distribution over 
equivalence-class representatives, giving rise to a huge loss
in the achievable rate compared to 
Figs.~\subreference{fig:rates}{fig:24_3}--\subreference{fig:rates}{fig:all}.   
Thus it is important that the transmitted symbol blocks are chosen
to be square-law distinct.

As shown in 
Figs.~\subreference{fig:rates}{fig:24_3}--\subreference{fig:rates}{fig:44rate},
the value of $\beta$ affects the minimum required power to achieve 
a target data rate, for a fixed constellation $\mathcal{K}$. 
We find that $\beta=0.9$ either outperforms other choices of $\beta$
at close-to-saturation data rates, or the loss is negligible compared 
to other values of $\beta$.  

Inspired by~\cite{ungerboeck}, Fig.~\subreference{fig:rates}{fig:all}
shows the mutual information for 
different constellations and with different values of $n$, all with 
$\beta=0.9$. This figure can be interpreted in two different ways, as follows.
First, by fixing the ROP at a constant value, we 
can achieve a higher data rate by choosing a larger constellation. For example,
at the ROP of $-16~$dBm and with $n=3$, 
a data rate of $2$~bit/sym is achievable with the
$2$-ring/$4$-ary phase constellation, while
a data rate of $2.7$~bit/sym is achievable  
with the $4$-ring/$4$-ary phase constellation. 
Secondly, by selecting a larger $\mathcal{K}$, 
a target data rate is achievable
at a smaller ROP. For example, the data rate of 
$2$~bit/sym is achievable by the $2$-ring/$4$-ary phase constellation
with $n=4$ at an ROP $\simeq-19~$dBm, while by using the $4$-ring/$4$-ary 
phase constellation, with $n=3$ and a channel code of rate
$\simeq 0.7$,
the same data rate is achievable at the ROP of $-24~$dBm, \textit{i.e.},
with approximately a $5~$dB gain. 
Note that the number of equivalence classes in the 
former is $432$, while in the latter it is $400$; therefore, their ML 
block-detection complexities are roughly the same.  

As discussed earlier in this section, 
for a fixed constellation $\mathcal{K}$, the achievable rate will 
increase with increasing $n$. This fact can be seen from 
Fig.~\subreference{fig:rates}{fig:all}. For example, 
while the maximum achievable rate for $4$-ring/$4$-ary phase constellation 
is $2.88$~bit/sym for $n=3$, the maximum achievable rate is 
$2.99$ and $3.08$~bit/sym for $n=4$ and $n=5$, respectively.
If one chooses the $90\%$ in-band energy as the criterion for defining the 
bandwidth, then the spectral efficiencies for $n=3$, $n=4$, and $n=5$ is, 
respectively, $2.94$, $3.05$, and $3.14$ bit/sec/Hz. 
Note that the maximum rate for this constellation is $4$ bit/sec/Hz under
coherent detection; therefore,
in the last two cases, \textit{i.e.}, $n=4$ and $n=5$, 
the rate is within the bounds
given in~\cite{capacity_bandlimited,capacity_timelimited}. 
Furthermore, although
the spectral efficiency for $n=3$ is below the rate lower-bound
\textit{i.e.}, $3$ bit/sec/Hz, the gap is small, \textit{i.e.},
about $2\%$. 
By the same bandwidth definition, 
the maximum spectral efficiency
for other constellations in Fig.~\subreference{fig:rates}{fig:all} 
are all within one bit/sec/Hz of coherent detection. 
However, if one chooses the $95\%$ in-band energy as the 
criterion for specifying the bandwidth then,
except the $4$-PSK constellation, the spectral-efficiency gap with
coherent detection is greater than one bit/sec/Hz. For example, 
the maximum spectral efficiency for $2$-ring/$4$-ary phase 
constellation with $n=4$ becomes $1.9$ bit/sec/Hz which is $5\%$
smaller than the lower bound, \textit{i.e.}, $2$ bit/sec/Hz. 

\begin{figure*}
\centering
\subfloat[$\mathcal{K}:2$-ring/$4$-ary phase constellation, $n=3$
	\label{fig:24_3}]{
	\includegraphics[scale=0.6666666]{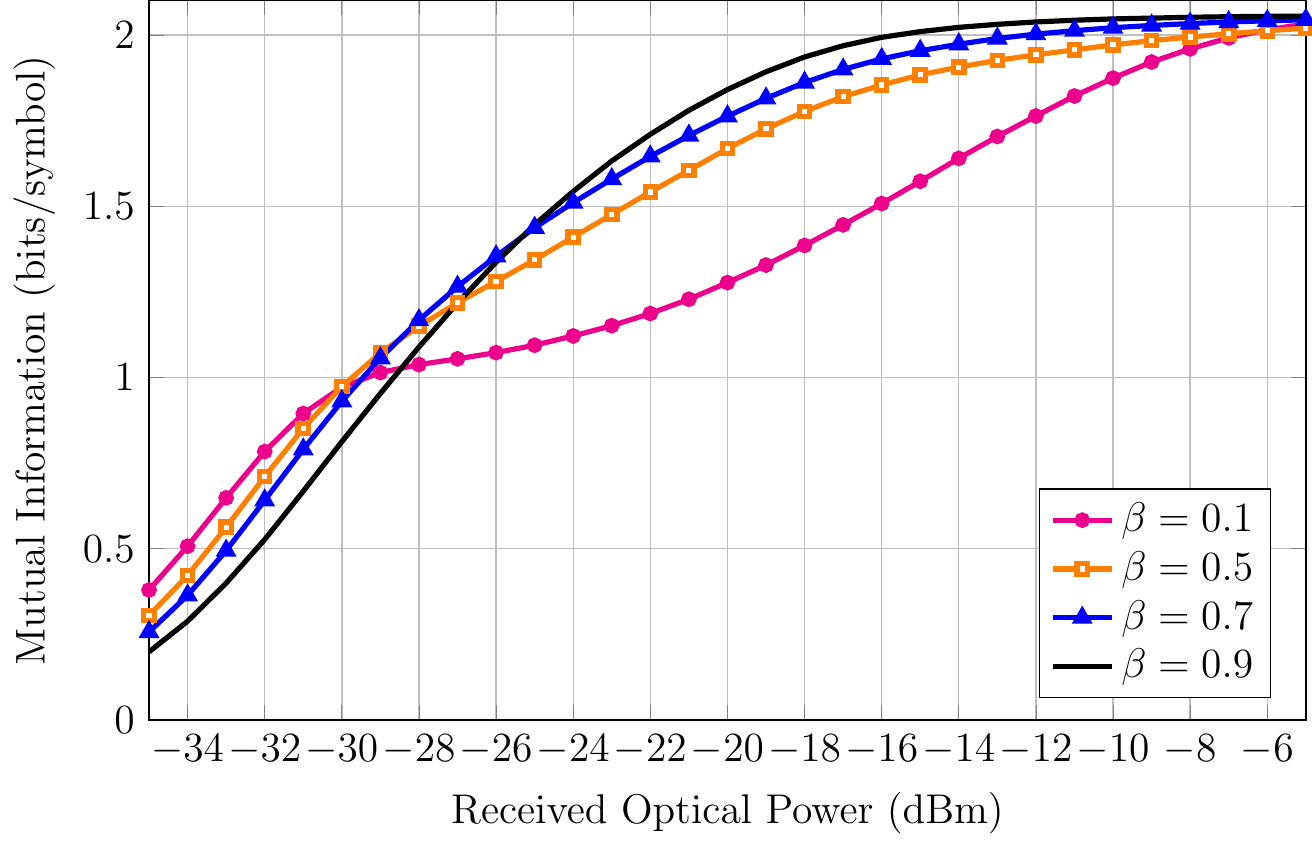}}
\subfloat[$\mathcal{K}:2$-ring/$4$-ary phase constellation, $n=4$
	\label{fig:24_4}]{
	\includegraphics[scale=0.6666666]{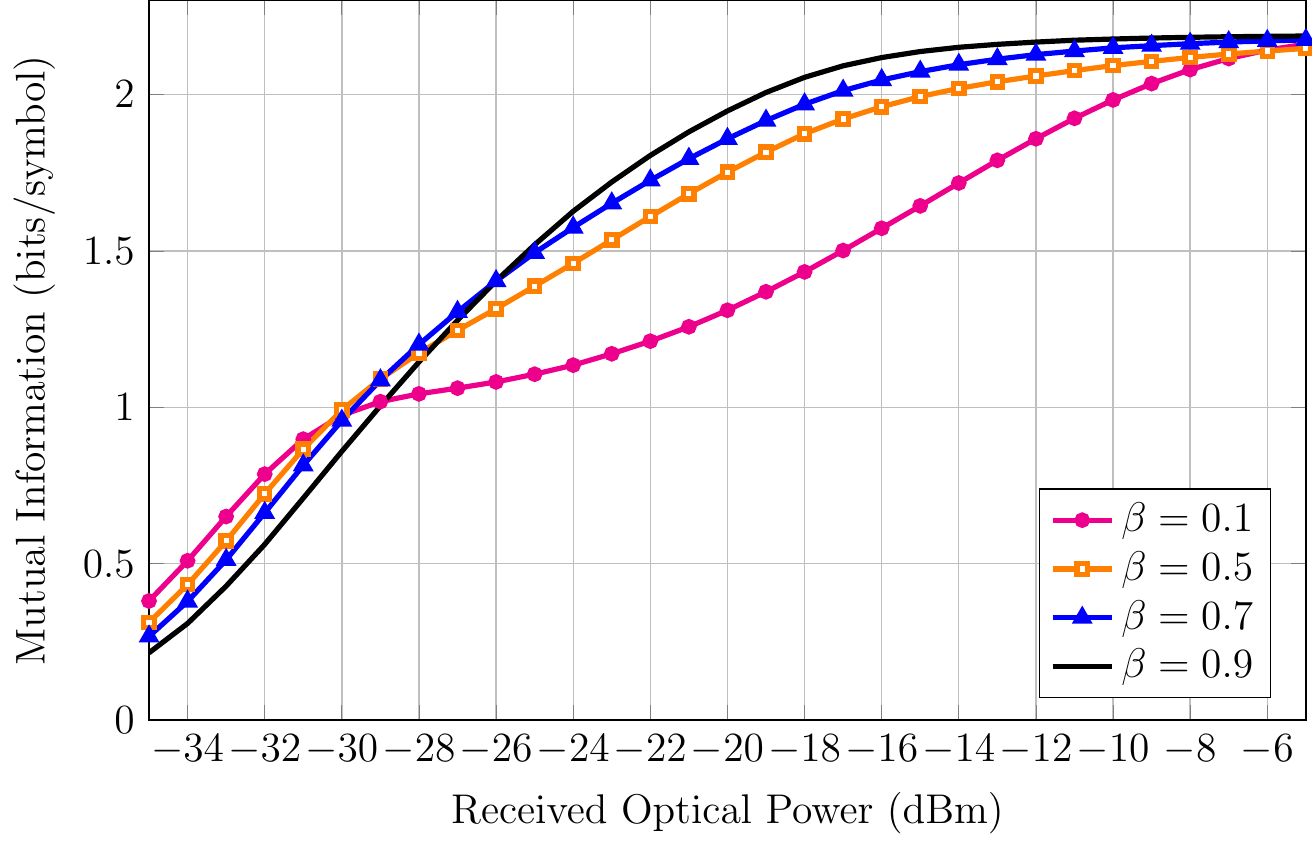}}\\
\subfloat[$\mathcal{K}:4$-PSK, $n=8$
	\label{fig:4pskrate}]{
	\includegraphics[scale=0.6666666]{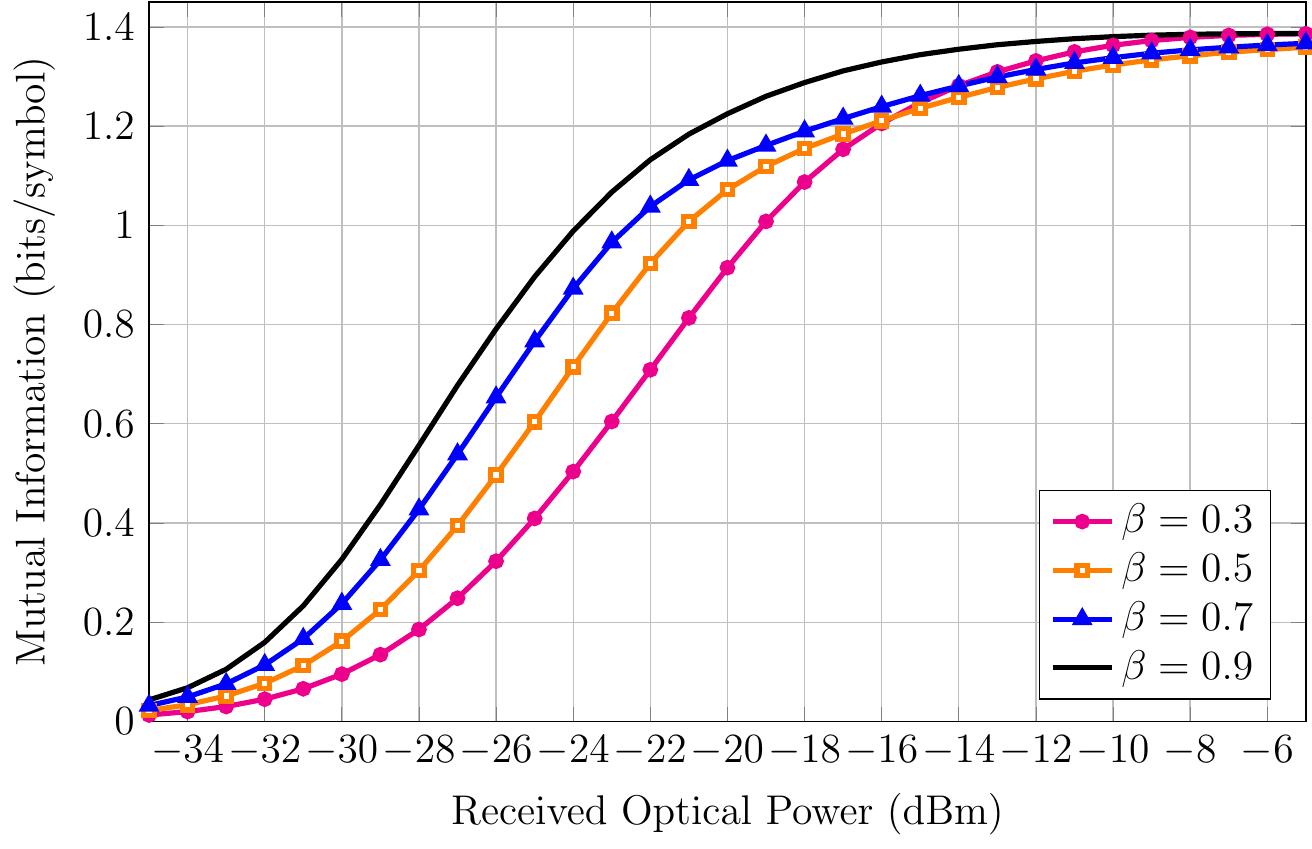}}
\subfloat[$\mathcal{K}:4$-ring/$4$-ary phase constellation, $n=3$
	\label{fig:44rate}]{
	\includegraphics[scale=0.6666666]{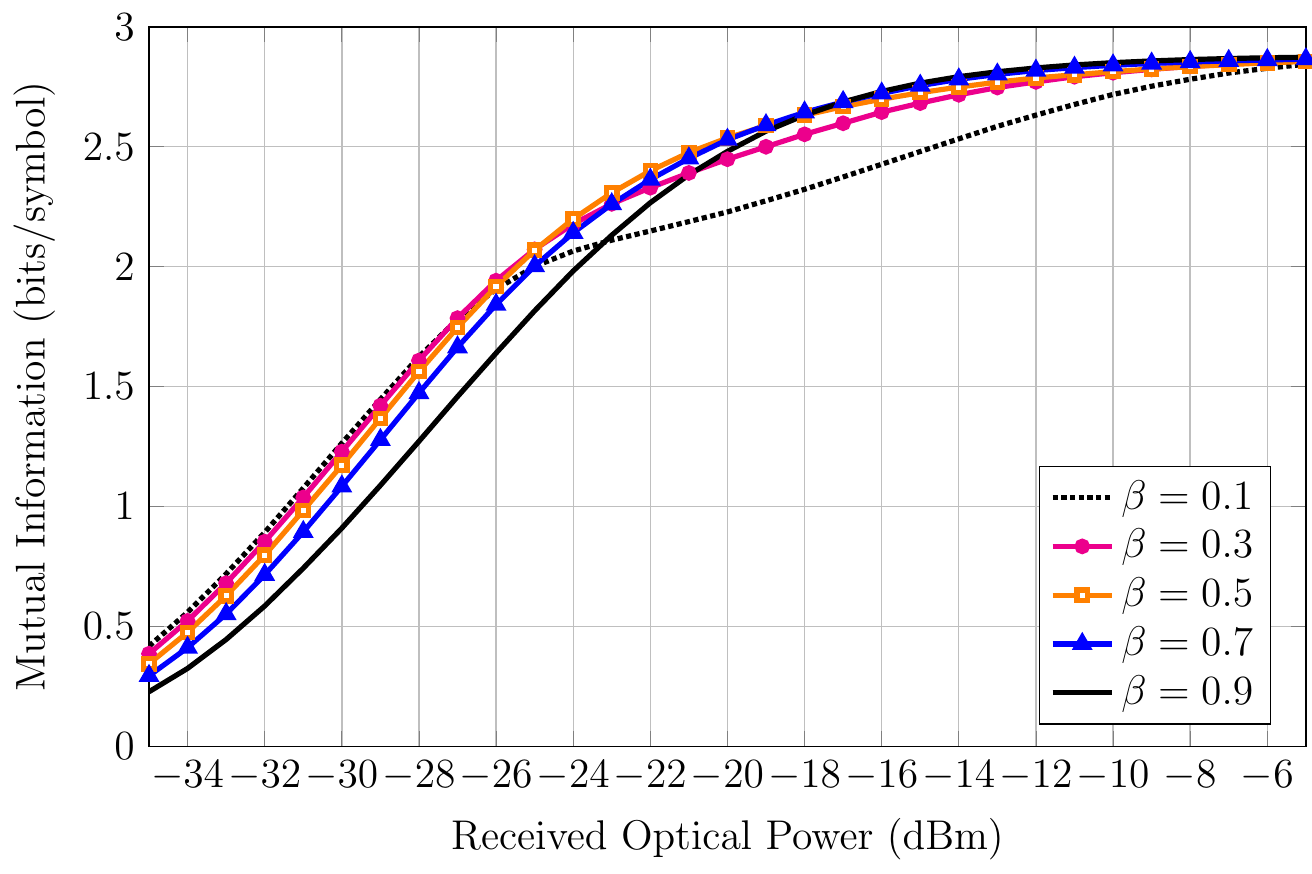}}\\
\subfloat[Uniform input distribution over all equivalence classes, $\beta=0.9$
	\label{fig:all}]{
	\includegraphics[scale=0.6666666]{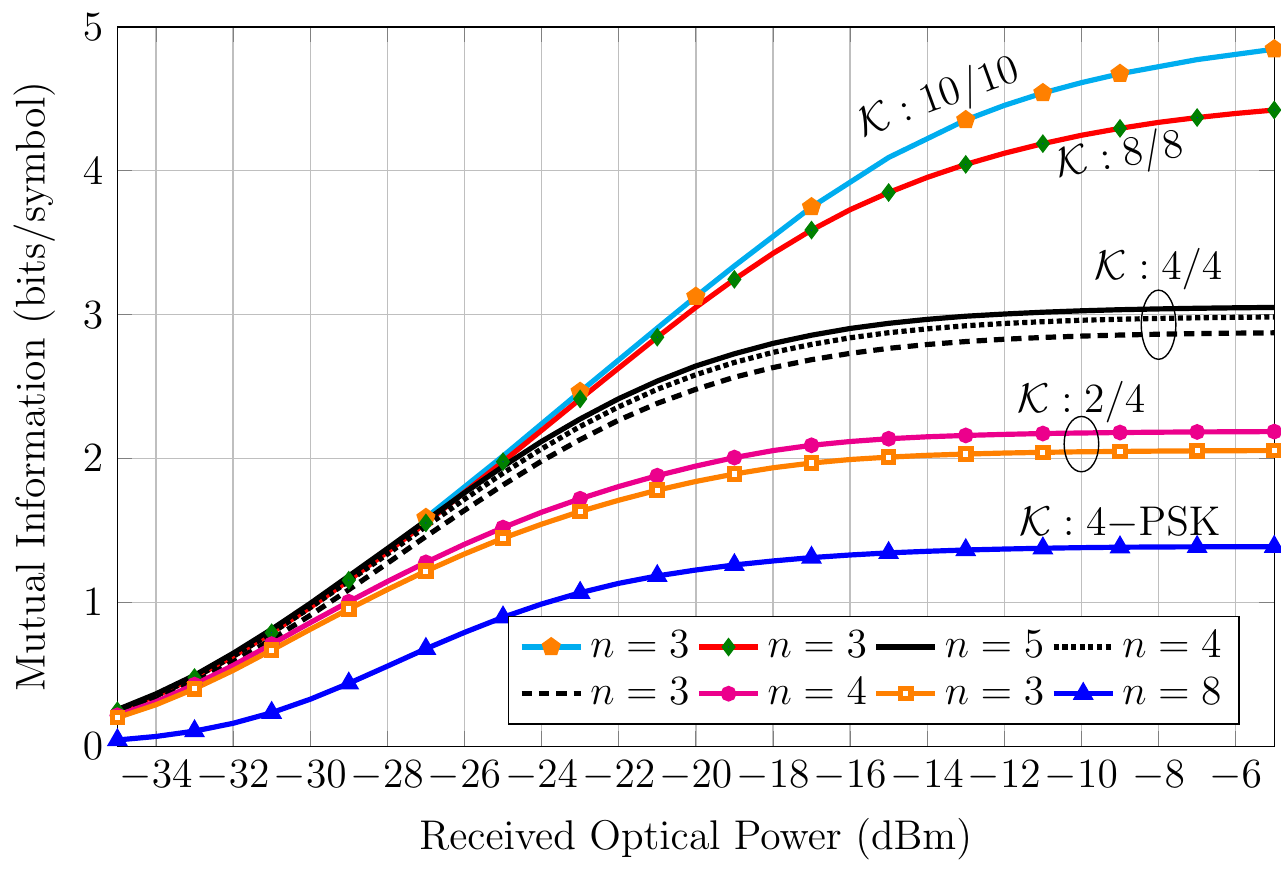}}
\subfloat[Uniform input distribution over $\mathcal{K}^4$, 
	$\mathcal{K}:2$-ring/$4$-ary phase constellation
	\label{fig:uniform}]{
	\includegraphics[scale=0.6666666]{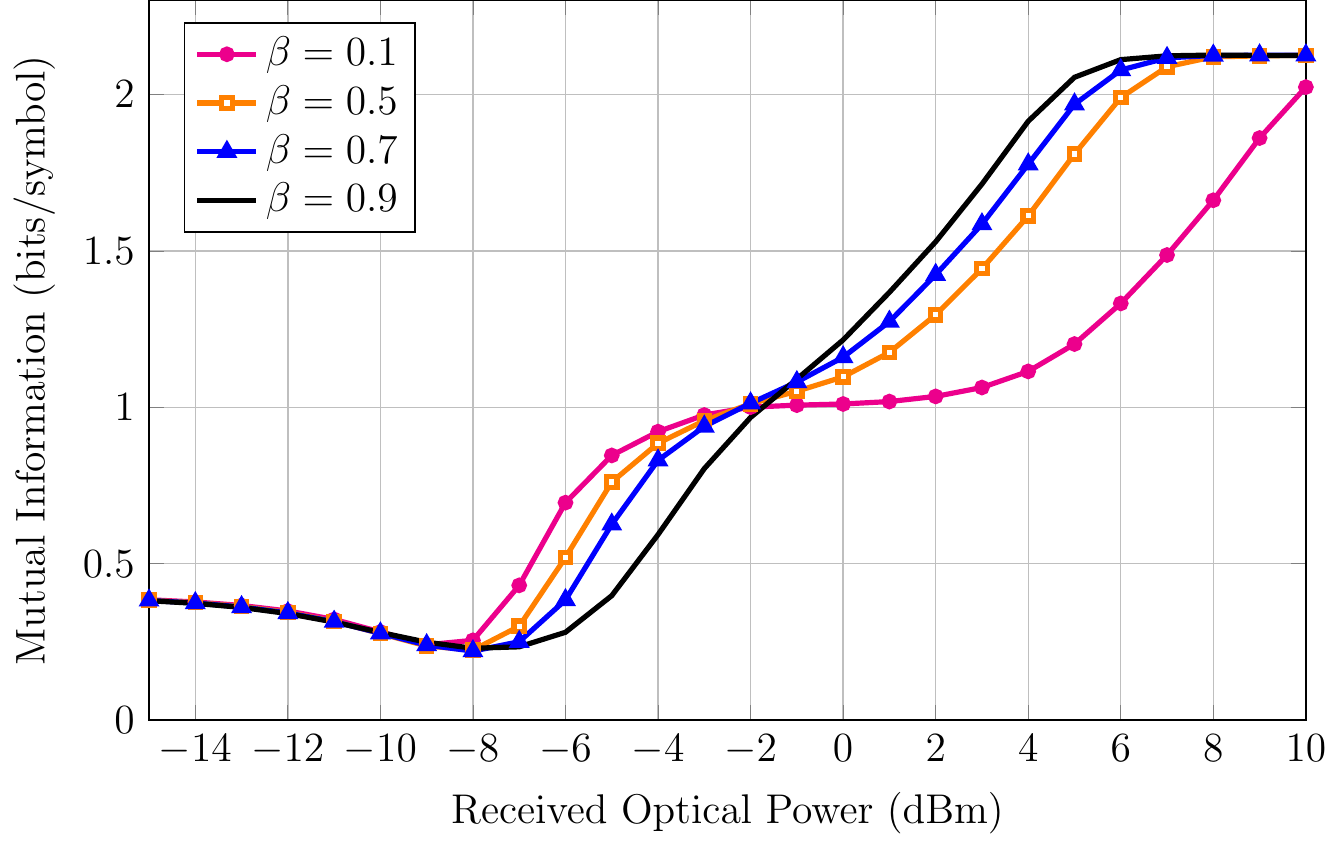}}
\caption{Mutual information in different scenarios. 
	In \protect\subref{fig:24_3}--\protect\subref{fig:all} the input 
	distribution is uniform over all equivalence classes, while 
	in \protect\subref{fig:uniform} the input distribution is uniform
	over $\mathcal{K}^4$, where $\mathcal{K}$ is the $2$-ring/4-ary phase
	constellation. In part \protect\subref{fig:all}, 
	$\mathcal{K}:a/b$ means that $\mathcal{K}$ is an 
	$a$-ring/$b$-ary phase constellation. Furthermore, the 
	maximum achievable rate for $8$-ring/$8$-ary and $10$-ring/$10$-ary phase 
	constellations, each with $n=3$, is $4.45$ and $4.96$ bits/symbol, respectively.}
\label{fig:rates}
\end{figure*}

\end{subsection}

\begin{subsection}{Bit Error Rate}
\label{subsec:ber}

Fig.~\ref{fig:ber} shows the bit error rate of the proposed scheme
in a back-to-back ($L=0$) configuration,
for several $\mathcal{K}$, $\beta$, and $n$. 
In each scenario, $M$ is chosen to be an integer power of $2$. 
For example, while there are $432$ equivalence classes for 
the $2$-ring/$4$-ary phase constellation with $n=4$, only $M=2^8$ 
of them are chosen as the symbol blocks, \textit{i.e.,} members of 
$\mathcal{S}$.  The length $\log_2(M)$ binary label associated with each
of the $M$ symbols was chosen randomly, i.e., no
particular labelling algorithm, \textit{e.g.}, 
Gray code, is being used. 
 
As shown in Fig.~\subreference{fig:ber}{fig:ber_24_small}, for the
$2$-ring/$4$-ary phase constellation, the BER decreases when
increasing $\beta$ from $0.3$ to $0.99$.  It might be perceived from
this figure that a larger value of $\beta$ results in a better BER
performance; therefore, the best BER performance is achieved for the
maximum possible value for $\beta$, \textit{i.e.}, unity.  However,
Fig.~\subreference{fig:ber}{fig:ber_24_large} refutes this claim.  As
shown in this figure, for close-to-unity values of $\beta$, a small
$\beta$ increment results in a significant performance loss, in terms
of BER. In the extreme case, \textit{i.e.}, $\beta=1$, there is no
ISI-free sample, and the $n$ complex-valued transmitted symbols must
be detected from the $n-1$ real-valued ISI-present received samples.
Thus, there is a significant performance loss, both in terms of BER
and the maximum achievable rate.

The BER performance under $4$-PSK and $4$-ring/$4$-ary phase
constellations are shown in Fig.~\subreference{fig:ber}{fig:ber_4psk}
and Fig.~\subreference{fig:ber}{fig:ber_44}, respectively. In both
cases, the performance is improved by increasing $\beta$ from $0.1$
toward $0.9$. 

Despite the fact that $\beta=0.99$ has the best BER performance in
Fig.~\subreference{fig:ber}{fig:ber_24_small} for the $2$-ring/$4$-ary
phase constellation, because of the discussed high sensitivity of
performance to the value of $\beta$ for $\beta>0.99$, $\beta=0.9$ is
used to compare the BER of different constellations in
Fig.~\subreference{fig:ber}{fig:ber_all}.  For the purpose of
fairness, the horizontal axis is normalized to information rate.
Similar to a typical communication over an additive white Gaussian
noise (AWGN) channel, for a fixed noise power and at high signal
powers, using a larger constellation results in a higher BER; however,
at low ROPs, the BER performance is a complicated function of the
received power.

The choice of constellation $\mathcal{K}$ has a big impact on the
performance of the proposed scheme. For example, while $16$-QAM
(quadrature amplitude modulation) is a typical constellation in
communication over AWGN channels, it has very poor performance in the
proposed scheme, as shown in
Fig.~\subreference{fig:ber}{fig:ber_16qam}.

The BER performance for a nonzero fiber length, \textit{i.e.}, $\rho<1$, 
for $2$-ring/$4$-ary and $4$-ring/$4$-ary phase constellations
are shown in Figs.~\subreference{fig:ber_ssfm}{fig:ber_24_ssfm} 
and~\subreference{fig:ber_ssfm}{fig:ber_44_ssfm}, 
respectively. 
The split-step Fourier method~\cite{ssfm}
was used to approximate wave propagation through a $10$~km 
standard single-mode fiber (SSMF) 
at an operating wavelength of $\lambda=1550~$nm.
One observes that, by using a $4$-ring/$4$-ary phase constellation with
$256$ symbol blocks of length $n=3$ and $\beta=0.9$, 
a BER of $10^{-3}$ is resulted at a 
$-10~$dBm launch power.  
As the power loss of SSMF at this wavelength is about 
$0.2~\text{dB}\cdot\text{km}^{-1}$; a $-10~$dBm launch power results
a $-12~$dBm ROP. Note that this agrees with 
Fig.~\subreference{fig:ber}{fig:ber_44},
\textit{i.e.}, a BER of $10^{-3}$ resulted from
same system parameters at a $-12~$dBm ROP. 
At higher launch powers, however,
there is a discrepancy
between the figures, resulting from
uncompensated nonlinear fiber impairments.
To achieve a high data rate, one may operate in the low-to-moderate
launch-power 
regime---where fiber nonlinearity is not 
a critical factor---by choosing high-order 
constellations. 
If high detection complexity rules high-order 
constellations out then one might use
low-order constellations and operate at high launch powers
by precompensating not only
the chromatic dispersion but also the
fiber nonlinear impairments at the transmitter. However, such
nonlinear precompensation, which might be implemented by a form
of digital back-propagation \cite{back_propagation},
will itself incur significant implementation complexity.
\begin{figure*}
\centering
\subfloat[$\mathcal{K}:2$-ring/$4$-ary phase constellation, $n=4$, $M=2^8$
	\label{fig:ber_24_small}]{
	\includegraphics[scale=0.6666666]{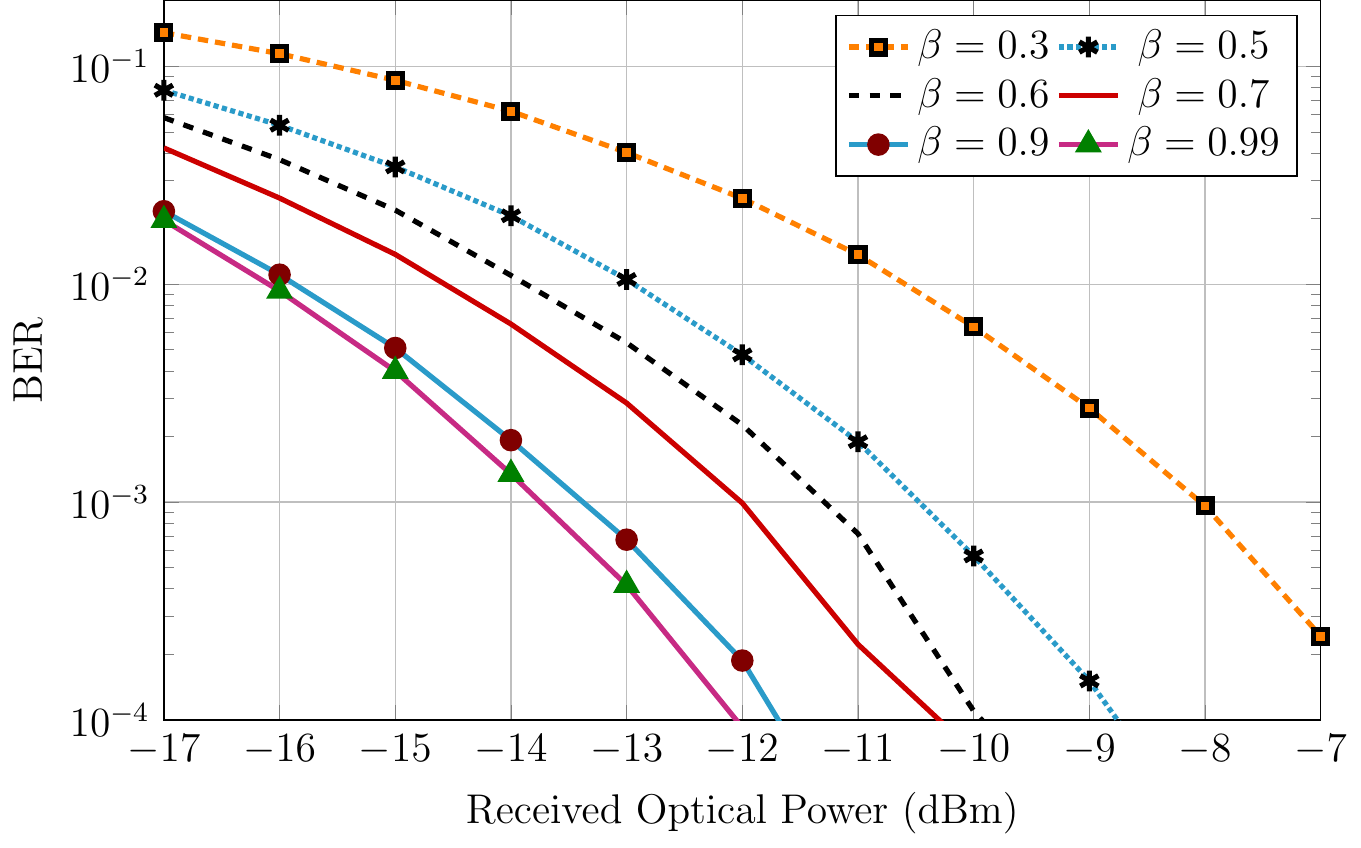}}
\subfloat[$\mathcal{K}:2$-ring/$4$-ary phase constellation, $n=4$, $M=2^8$
	\label{fig:ber_24_large}]{
	\includegraphics[scale=0.6666666]{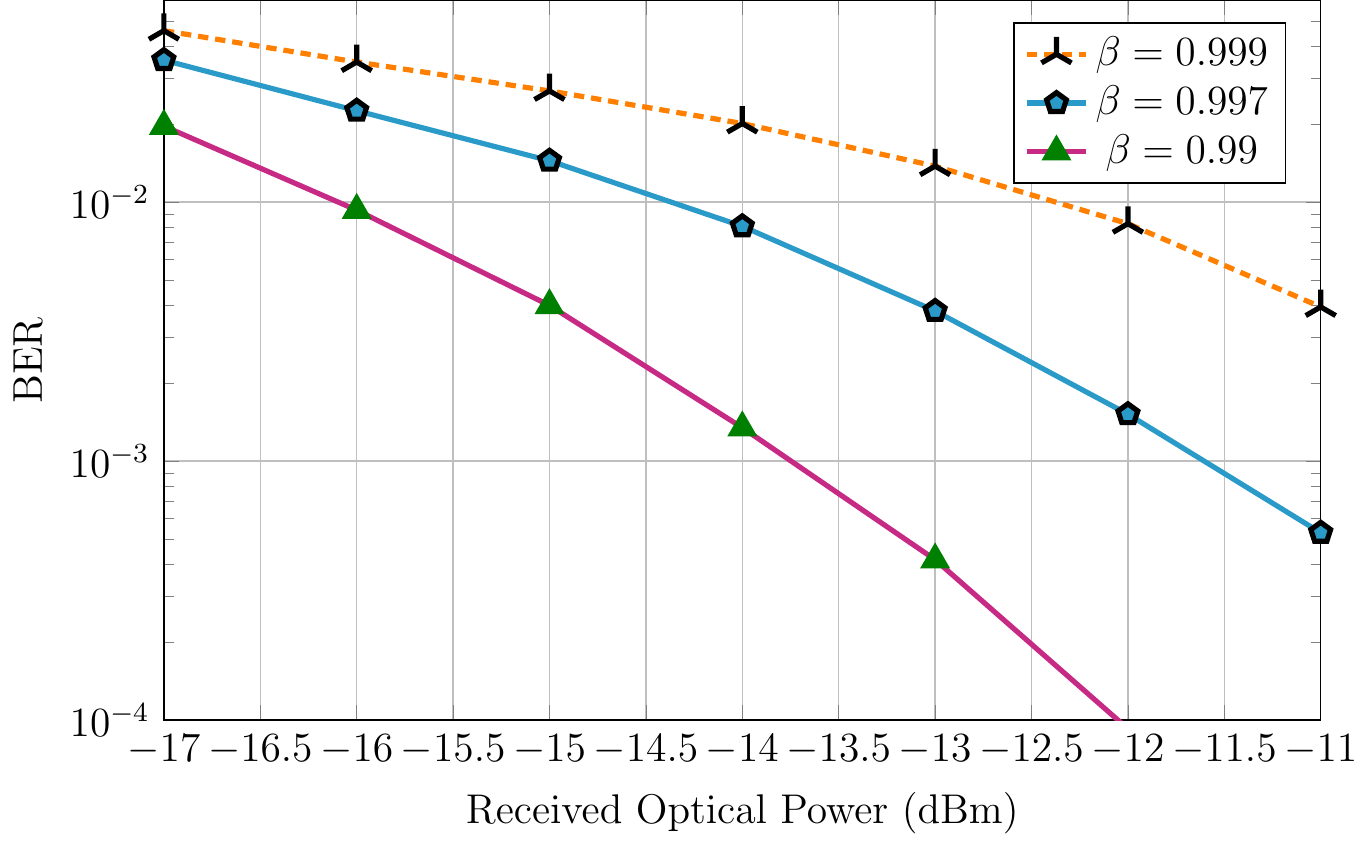}}\\
\subfloat[$\mathcal{K}:4$-PSK, $n=8$, $M=2^{11}$
	\label{fig:ber_4psk}]{
	\includegraphics[scale=0.6666666]{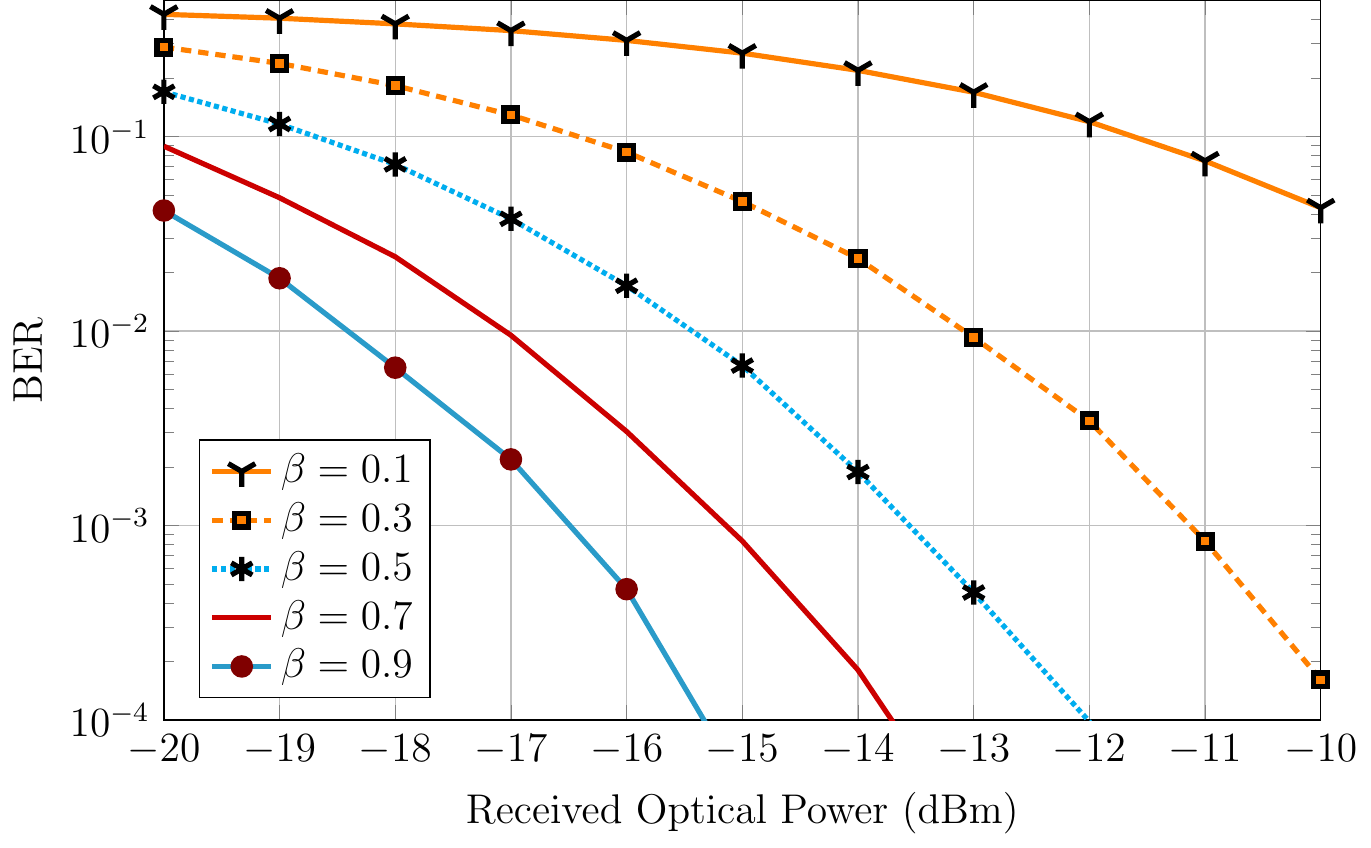}}
\subfloat[$\mathcal{K}:4$-ring/$4$-ary phase constellation, $n=3$, $M=2^8$
	\label{fig:ber_44}]{
	\includegraphics[scale=0.6666666]{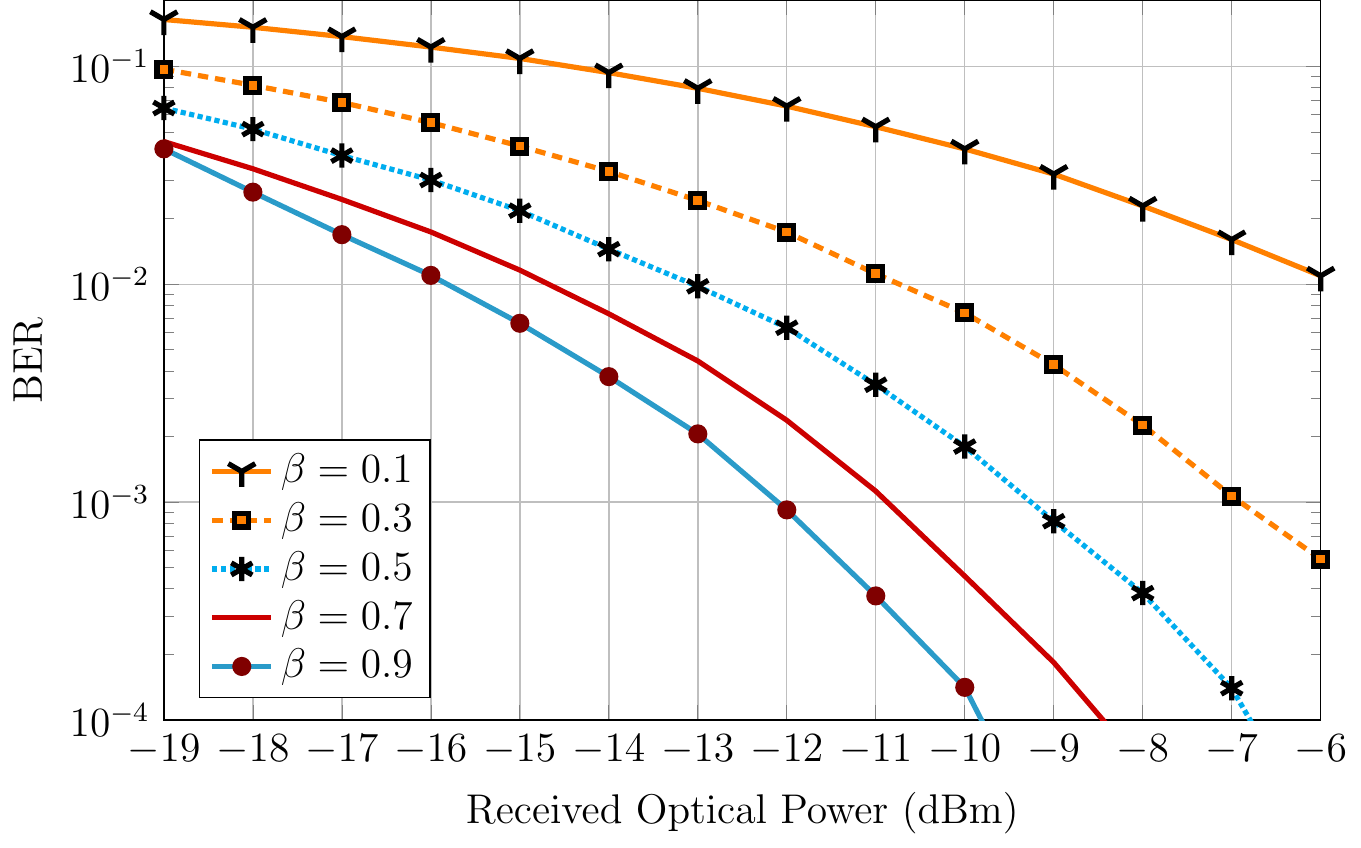}}\\
\subfloat[Different constellations with $\beta=0.9$
	\label{fig:ber_all}]{
	\includegraphics[scale=0.6666666]{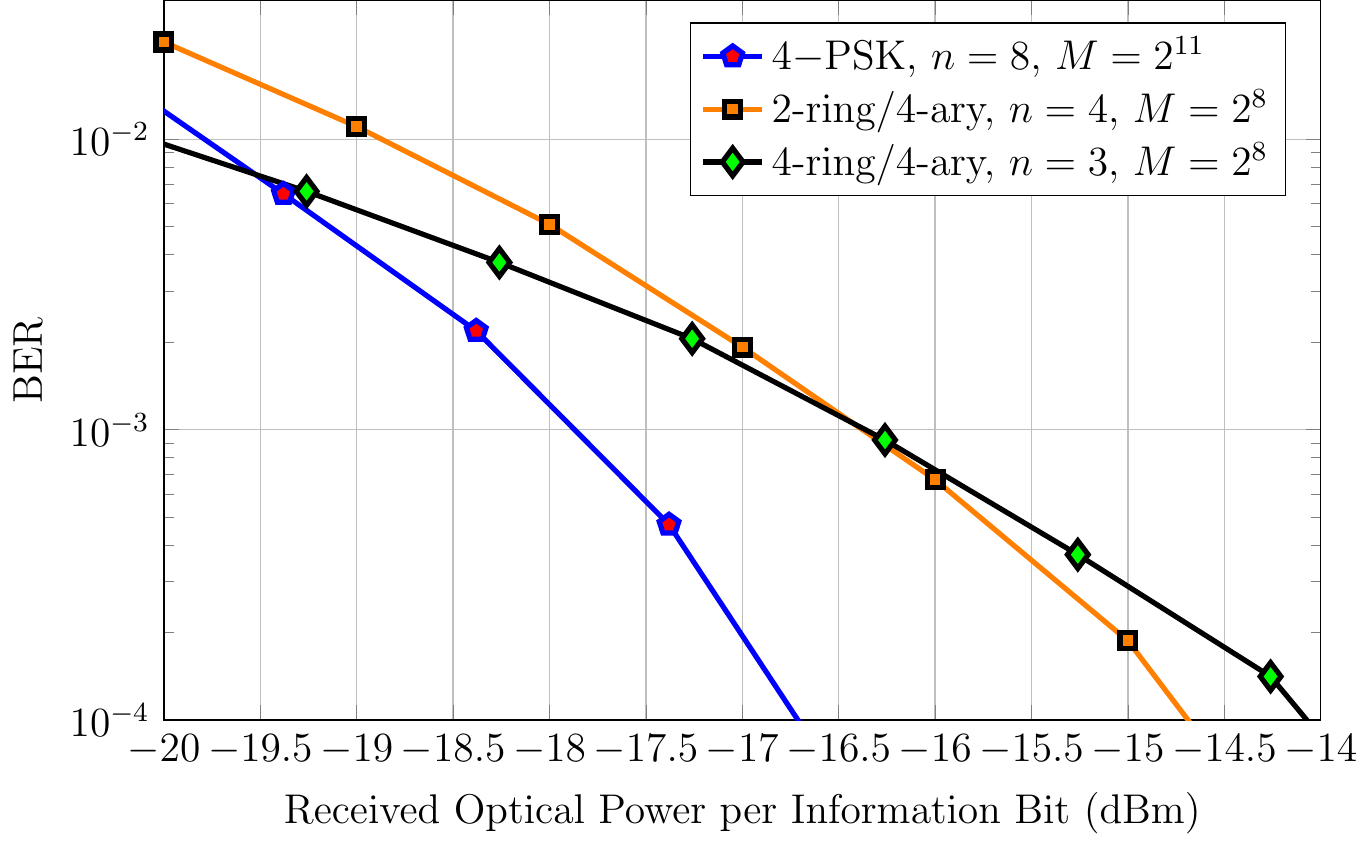}}
\subfloat[$\mathcal{K}:16$-QAM, $n=3$, $M=2^8$
	\label{fig:ber_16qam}]{
	\includegraphics[scale=0.6666666]{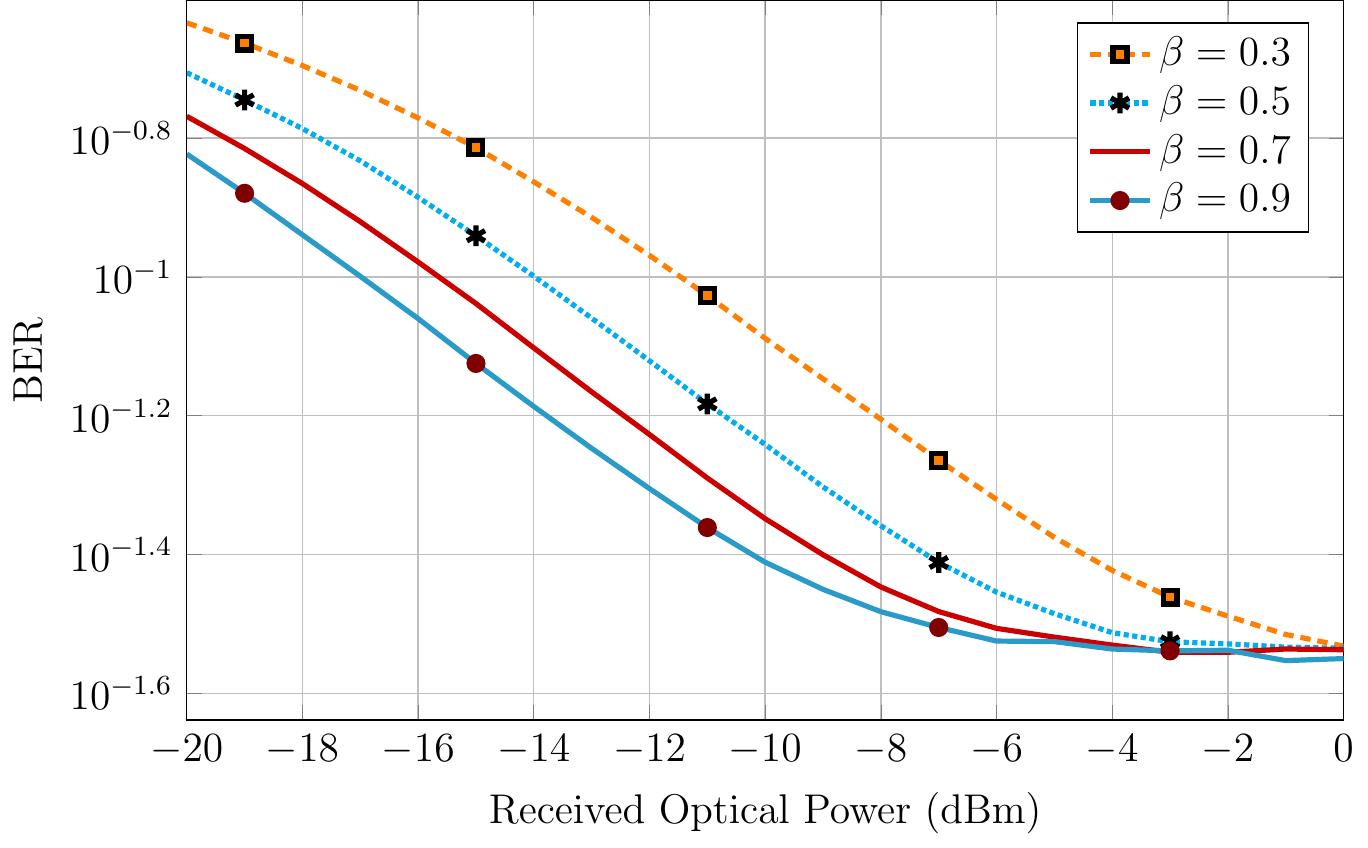}}\\
\caption{Bit error rate in back-to-back transmission in different scenarios.
	Part \protect\subref{fig:ber_16qam} 
	suggests that the conventional $16$-QAM constellation does
	not have a good performance under the proposed communication scheme.}
\label{fig:ber}
\end{figure*}

\begin{figure}
\centering
\subfloat[$\mathcal{K}:2$-ring/$4$-ary phase constellation, $n=4$, $M=2^8$
	\label{fig:ber_24_ssfm}]{\centering
	\includegraphics[scale=0.6666666]{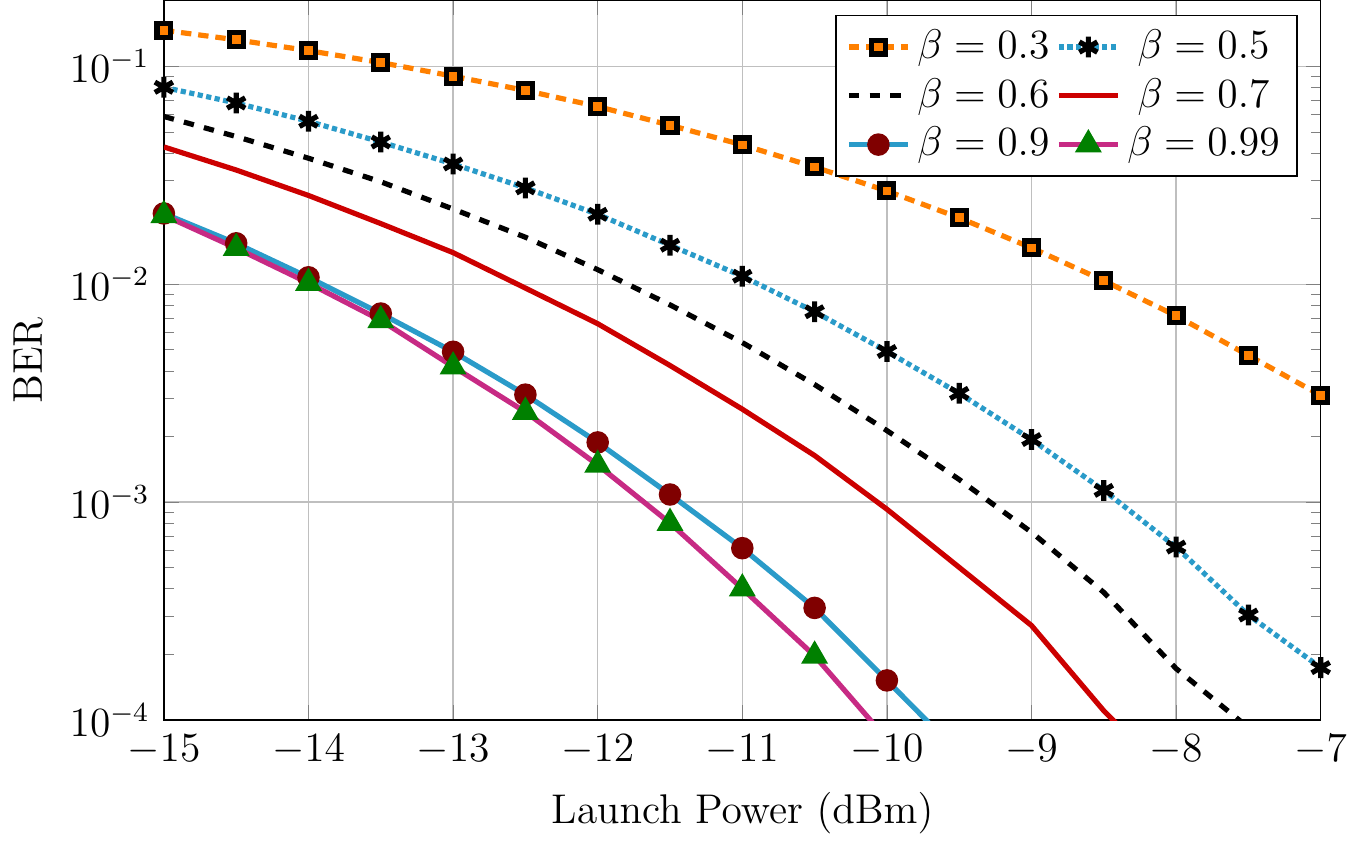}}
	\hfil
\subfloat[$\mathcal{K}:4$-ring/$4$-ary phase constellation, $n=3$, $M=2^8$
	\label{fig:ber_44_ssfm}]{\centering
	\includegraphics[scale=0.6666666]{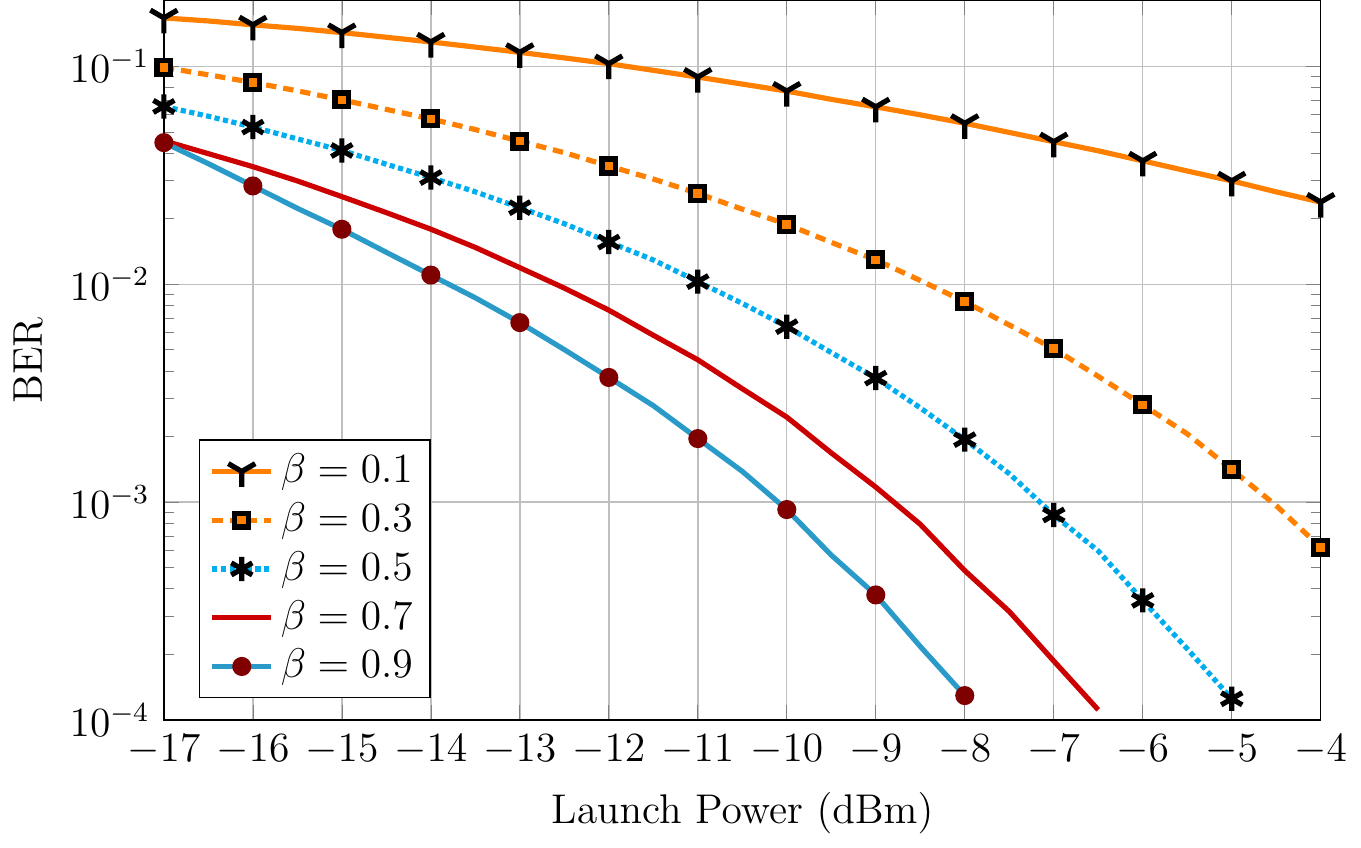}}
	\hfil
\caption{BER performance of the proposed scheme 
	for a $10~$km standard single-mode fiber at the $1550$~nm operating
	wavelength. The propagation through the optical
	fiber is approximated by the split-step Fourier method.}
\label{fig:ber_ssfm}
\end{figure}
\end{subsection}

\begin{subsection}{Power}
\label{subsec:psd}

If the transmitted symbols are i.i.d., 
Theorem~\ref{theorem} guarantees that the average power of the
transmitted waveform
is equal, in probability, to the variance of those symbols. On the
other hand, in 
Sec.~\ref{subsec:mutual_information} it was noted that 
uniform selection of symbol blocks from the entire $\mathcal{K}^n$
results in a significant loss in achievable rate,
and therefore must be avoided. 
Constraining symbol blocks to be square-law distinct
makes the transmitted symbols dependent, and thus
Theorem~\ref{theorem} does not apply.
However,
in the numerical simulations
it turned out that 
the average power of 
the transmitted waveform is indeed equal to $\mathcal{P}_\mathcal{S}$
in all cases studied, where
$\mathcal{P}_\mathcal{S}\triangleq\frac{1}{Mn}
	\sum_{j=1}^{M}\lVert\bm{x}_j\rVert^2$
is the average power of $\mathcal{S}$
and where $\lVert\cdot\rVert$ denotes the $L^2$-norm.

\end{subsection}

\begin{subsection}{Comparison With Other Schemes}
\label{subsec:compare}

Due to differences in channel models,
particularly the idealization of optical fibers,
a fair comparison of the
proposed scheme with other existing schemes is difficult. 
For example,
while we have assumed that there are no optical amplifiers, and, as a result,
no ASE noise in the channel, 
many papers in the literature use
erbium-doped fiber amplifiers, which are primary sources of ASE noise.

Although not exactly fair,
to make a comparison, let us assume the proposed scheme is used to transmit
data over a span of $10~$km of SSMF. The resulting power loss at $1550$~nm is
$2~$dB.   Factoring in an
additional $5~$dB loss due,
\emph{e.g.}, to connector losses and code operation at some gap to the Shannon
limit, let us
assume the launch power is $7~$dB larger than the received power.
In this case,
from Fig.~\subreference{fig:rates}{fig:all} we see that 
an information rate of 
$4~$bits per symbol can be achieved with the proposed scheme, 
either by using the $8$-ring/$8$-ary phase
constellation with $n=3$ and a channel code of rate $9/10$
at a launch power of $-6.5~$dBm, or by using 
the $10$-ring/$10$-ary phase constellation with $n=3$ and a channel code
of rate $8/10$ at a launch power of $-8.5~$dBm.  
On the other hand,
using a Kramers--Kronig receiver,
the authors of~\cite{kk6}
achieve a data rate of $4$~bits per symbol
at a launch power of $2~$dBm.  
However, as noted, the channel model of \cite{kk6} is quite different.
\end{subsection}

\end{section}
\begin{section}{Complexity}
\label{sec:complexity}

In this section we discuss the implementation
complexity of direct detection under
Tukey signalling.  Note that the proposed scheme aims to achieve
spectral efficiencies close to those of a coherent
detector. Therefore---instead of comparing with simple IM/DD schemes
whose rates are roughly half the data rate under coherent
detection---the complexity of the proposed scheme must be compared
with schemes that modulate complex-valued waveforms, \textit{e.g.},
coherent and Kramers--Kronig transceivers.

Fig.~\ref{fig:id} shows simple examples of
analog and digital integrate-and-dump
circuits.  Other circuit topologies,
designed specifically for high baud rate
fiber-optic communications, are well investigated~\cite{id}.  In
order to integrate the output of the photodiode over the ISI-free and
ISI-present intervals, a clock with a duty cycle of $\beta$ and its
inverse is needed. Two analog-to-digital converters (ADCs),
each operating at the baud rate, are needed to convert
$s(\cdot)$ into discrete-time output samples, \textit{i.e.}, $y_k$'s
and $z_\ell$'s, where $k\in\{0,\ldots,n-1\}$ and
$\ell\in\{0,\ldots,n-2\}$.  This number of ADCs is exactly the same
as required (per polarization) in coherent detection, 
as shown in Fig.~\ref{fig:compare}.
The spectrum-broadening operations in
typical Kramers--Kronig receivers necessitates even higher conversion rates,
\textit{e.g.,} six times the baud rate. However, there are some
schemes that allow sampling at twice the baud rate, but under some
assumptions, \textit{e.g.}, high carrier-to-signal power
ratio~\cite{kk9,kk10}.

Similar to coherent and Kramers--Kronig transceiver schemes, both the in-phase
and the quadrature components of the transmitted waveform are modulated.  In
other words, the electric field itself (and not just the field intensity) is
modulated.  Therefore, an \textit{IQ modulator} 
is needed at the transmitter. 

As mentioned in Sec.~\ref{subsec:tukey_signalling}, 
$w_\beta(\cdot)=h_\beta(\cdot)\ast\Pi(\cdot)$; this 
property can be exploited in implementing the Tukey waveform at the transmitter.
In fact,
two zero-order-hold digital-to-analog converters (DACs), 
one for the in-phase and one for the quadrature components,
each operating at the baud rate and of hold-duration $T$, 
can be used to map the discrete-time transmitted symbols 
into a train of rectangular waveforms which itself is then 
filtered by an LTI filter with impulse response
$h_\beta\left(\frac{t}{T}\right)$,
producing $x(\cdot)$. Since $h_\beta(\cdot)$ has bounded support, 
the filter has a causal implementation.

Equivalence
class representatives for a particular constellation $\mathcal{K}$ and block
length $n$ can be pre-computed and stored in a look-up table.  Thus there
is no operational complexity related to determining the set $\mathcal{S}$.

The naive scheme for ML block detection searches over all $|\mathcal{S}|$
transmitted blocks to maximize the likelihood score. Therefore, its complexity
is $\mathcal{O}(2^{nr_\text{max}})$, where $r_\text{max}$ is the maximum
achievable rate with $\mathcal{S}$. As mentioned in
Sec.~\ref{subsec:transmitted_symbols}, a small $n$ suffices to achieve a target
BER; thus, the complexity of the ML block detection is a relatively small
fixed constant.
It may be possible to reduce this complexity, for example,
via a trellis-search algorithm, but we leave the investigation of this
for future work.

\newsavebox{\idbox}
\begin{figure}
\centering
\sbox{\idbox}{\includegraphics[scale=1]{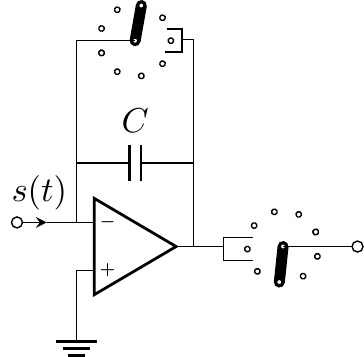}}
\subfloat[Analog circuit\label{fig:id_analog}]{\usebox{\idbox}}\hspace*{-1.5cm}
\subfloat[Digital circuit\label{fig:id_digital}]{\vbox to \ht\idbox{%
\vfil
	\includegraphics[scale=1]{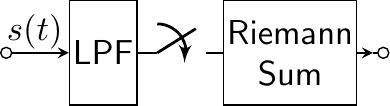}
\vfil}}
\caption{Simple integrate \& dump circuits, 
LPF: low-pass filter. While the forked wiring 
above the capacitor in (a) 
short-circuits the capacitor at the start of each 
integration interval, the one in front of the 
op-amp samples the integrator output. 
Improved integrate-and-dump schemes can be found in~\cite{id}.
}
\label{fig:id}
\end{figure}
\begin{figure}
\centering
	\subfloat[Intensity modulation with direct detection (IM/DD)\label{fig:comp_imdd}]{
	\includegraphics[scale=0.6666666666]{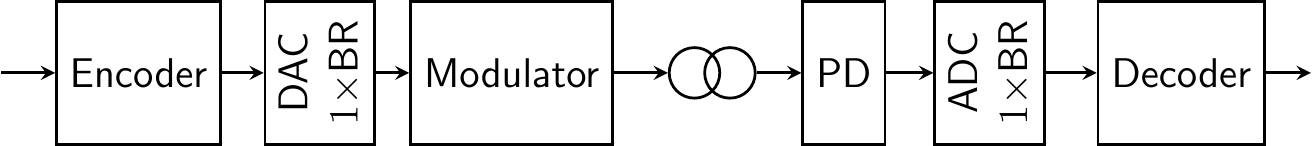}
	}\\
	\subfloat[Single-polarization coherent detection\label{fig:comp_coh}]{
	\includegraphics[scale=0.6666666666]{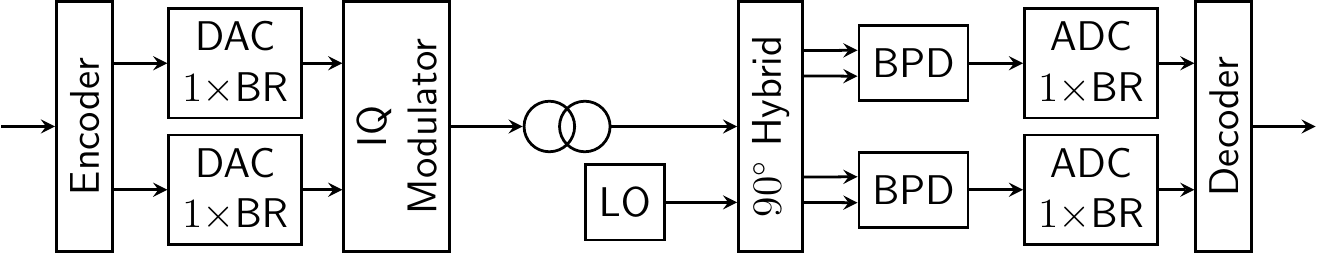}
	}\\
	\subfloat[Kramers--Kronig receiver\label{fig:comp_kk}]{
	\includegraphics[scale=0.6666666666]{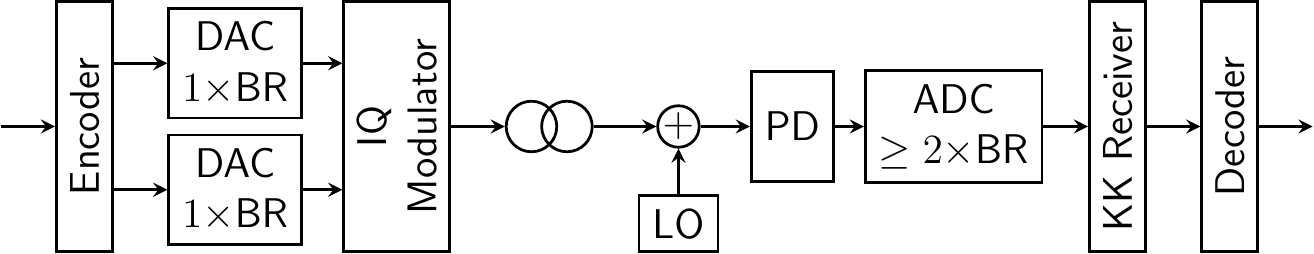}
	}\\
	\subfloat[The proposed scheme\label{fig:comp_tukey}]{
	\includegraphics[scale=0.6666666666]{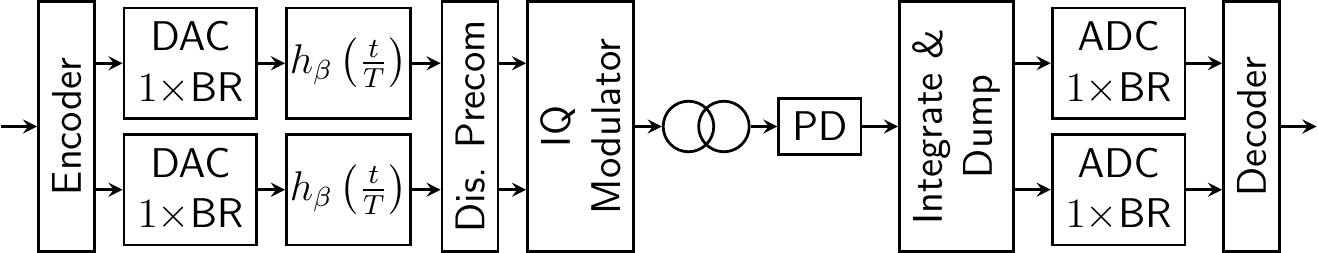}
	}
\caption{Abstract block diagrams for four different single-polarization transceiver schemes. 
	BR: baud rate,
	DAC: digital to analog converter, 
	ADC:analog to digital converter, 
	PD: photodiode, 
	BPD: balanced photo detector, 
	LO:local oscillator, 
	Dis. Precom: dispersion precompensation.
	The numbers below DACs and ADCs show the conversion speed compared to the baud rate.
	In the Kramers--Kronig receiver, instead of the receiver optical front-end,
	the tone can be injected to the signal at the transmitter back-end.
	DACs in (d) are of zero-order-hold type.	
	Except the IM/DD scheme, the modulated waveform is complex valued in the other ones.}
\label{fig:compare}
\end{figure}
\end{section}

\begin{section}{Discussion}
\label{sec:discussion}

As well stated by Tukey, 
``the test of a good procedure is how well it works, not how well it is
understood''~\cite{tukey5}.  So far
the proposed scheme has been tested only with computer simulations. It would be 
interesting to investigate its performance in a practical experiment.

Throughout this paper, we have assumed an unamplified optical link and
therefore no ASE noise in the system. It would be interesting to investigate
the performance of the proposed scheme in the presence of ASE noise.

We have used maximum-likelihood block detection to detect the transmitted
symbol block. It would be interesting to investigate other detection
strategies.  For example, we have assumed that blocks are transmitted without
guard time, which means that the last symbol of one block interferes with the
first symbol of the next block.  While our block detection rule ignores the
signal received during these intervals of overlap, a sequence-oriented
detection algorithm could potentially exploit them.

Due to chromatic dispersion, polarization mode dispersion, and nonlinear interactions between signals, the optical fiber channel is not memoryless.
In our proposed system we have precompensated for chromatic dispersion (a dominant cause of temporal signal broadening, and hence memory); however,
we have not studied or compensated for other impairments causing
memory.  Although
these factors may not be significant at
short transmission lengths or with relatively low launch power, it would
interesting to investigate the transmission
regimes where these factors begin to impact performance.

As noted in Sec.~\ref{sec:ml}, unlike AWGN channels, Euclidean distance is not
an appropriate metric for detecting the transmitted symbol block.  It would be
interesting to derive an appropriate metric for the proposed scheme.  In
addition to giving an insight on detection, such a metric might simplify the
detector's implementation. 

We have used only five different constellations to derive the results in 
this paper, namely, the $2$-ring/$4$-ary phase, the $4$-ring/$4$-ary phase, 
and the $4$-PSK constellations, and we have shown that $16$-QAM 
gives poor performance. 
Constellation design is thus
another interesting research topic that can be addressed in
future work.
In particular, given a set of performance criteria,
it would be interesting to
find the ``best'' constellation that satisfies the conditions. 
This question is closely related to the question about metrics,
since for a given power,
the performance of a constellation improves by increasing the 
minimum distance between its points.


We did not impose any criterion for choosing equivalence-class representatives,
\textit{i.e.}, symbol blocks.  It would be interesting to see if some
particular elements are better than others for use as class
representatives.

Although the numerical simulations revealed that the average transmitter 
power equals $\mathcal{P}_\mathcal{S}$ in the simulations,
it would be interesting to generalize Theorem~\ref{theorem}
to allow for dependent symbols. Strengthening the convergence type
is another relevant problem.

No doubt many further interesting problems can be posed.

\end{section}

\appendix[Proof of Theorem~\ref{theorem}]
For $k\in\{0,\ldots,m-1\}$ and $\ell\in\{0,\ldots,m-2\}$, let
the ISI-free interval $\mathcal{Y}_k$ and the ISI-present interval 
$\mathcal{Z}_\ell$ be as given in (\ref{eq:interval_y}) and 
(\ref{eq:interval_z}), respectively. Then,
\begin{equation}
	\frac{1}{T}\int_{-\infty}^{\infty}|\Lambda_m(t)|^2\text{d} t=
\sum_{k=0}^{m-1}G_k+\sum_{\ell=0}^{m-2}H_\ell+\zeta_m,
\label{eq:apx1}
\end{equation}
where $G_k\triangleq\frac{1}{T}\int_{\mathcal{Y}_k}|\Lambda_m(t)|^2\text{d} t$, 
$H_\ell\triangleq\frac{1}{T}\int_{\mathcal{Z}_\ell}|\Lambda_m(t)|^2\text{d} t$,
and 
\begin{equation*}
	\zeta_m=\frac{1}{T}\int_{-\infty}^{0}|\Lambda_m(t)|^2\text{d} t+\frac{1}{T}
	\int_{\left(m-\frac{1+\beta}{2}\right)T}^{\infty}|\Lambda_m(t)|^2\text{d} t.
\end{equation*}
One may see that 
$\expect{G_k}=\frac{4(1-\beta)}{4-\beta}P$
and
$\expect{H_\ell}=\frac{3\beta}{4-\beta}P$.

Let $\overline{G}_m$ and $\overline{H}_{m-1}$ denote, respectively,
the sample average of 
$G_k$ and $H_\ell$ random variables, \textit{i.e.},
$\overline{G}_m=\frac{1}{m}\sum_{k=0}^{m-1}G_k$
and
$\overline{H}_{m-1}=\frac{1}{m-1}\sum_{\ell=0}^{m-2}H_\ell$.
Then, (\ref{eq:apx1}) implies that
\begin{equation}
	\frac{1}{mT}\int_{-\infty}^{\infty}|\Lambda_m(t)|^2\text{d} t=
\overline{G}_m+\frac{m-1}{m}\overline{H}_{m-1}+\frac{\zeta_m}{m}.
\label{eq:equality_apendix}
\end{equation}
As $\lambda_k$'s are i.i.d. for 
$k\in\{0,\ldots,m-1\}$, the $G_k$ random variables are
i.i.d. and have finite variance, where the latter is true 
as $\var{|\lambda_k|^2}<\infty$.
As a result, 
by the weak law of large numbers, 
\begin{equation}
\overline{G}_m\overset{p}{\rightarrow}\frac{4(1-\beta)}{4-\beta}P
\quad\text{as }m\rightarrow\infty.
\label{eq:convergence_g}
\end{equation}
Unlike $G_k$'s, the $H_\ell$ random variables, $\ell\in\{0,\ldots,m-2\}$, 
are dependent. 
However, the dependency is only between adjacent $H_\ell$'s.
In particular, 
$H_i$ and $H_j$ are dependent,
thus, $\cov{H_i}{H_j}>0$, 
only if $|i-j|\leq 1$, where $i$ and $j\in\{0,\ldots,m-2\}$ and 
$\cov{\cdot}{\cdot}$ denotes the covariance. 
Furthermore, 
as $\var{|\lambda_k|^2}<\infty$ for all $k\in\{0,\ldots,m-1\}$,
it implies that
$\cov{H_i}{H_j}<\infty$. 
Therefore, 
\begin{equation*}
\var{\overline{H}_{m-1}}=\frac{\sum_{i,j=0}^{m-2}\cov{H_i}{H_j}}{(m-1)^2}=
\mathcal{O}\left(\frac{1}{m}\right).
\end{equation*}
Therefore, $\lim_{m\rightarrow\infty}\var{\overline{H}_{m-1}}=0$,
and Chebyshev's inequality implies that 
\begin{equation}
\overline{H}_{m-1}\overset{p}{\rightarrow}\frac{3\beta}{4-\beta}P
\quad\text{as } m\rightarrow\infty.
\label{eq:convergence_h}
\end{equation}
Furthermore, $\var{\frac{\zeta_m}{m}}=\mathcal{O}\left(\frac{1}{m^2}\right)$;
thus, 
\begin{equation}
\frac{\zeta_m}{m}\overset{p}{\rightarrow}0\quad\text{as }m\rightarrow\infty.
\label{eq:convergence_last} 
\end{equation}
As a result, (\ref{eq:equality_apendix}), (\ref{eq:convergence_g}),
(\ref{eq:convergence_h}), and (\ref{eq:convergence_last}) imply that
\begin{equation*}
	\frac{1}{mT}\int_{-\infty}^{\infty}|\Lambda_m(t)|^2\text{d} t\overset{p}{\rightarrow}
P\quad\text{as }m\rightarrow\infty.
\end{equation*}

\section*{Acknowledgment}
The authors would like to thank Qun Zhang for helpful discussions
regarding the split-step Fourier simulation method and an anonymous reviewer
whose comments significantly improved the paper.

\IEEEtriggeratref{46} 

\bibliographystyle{IEEEtran}
\bibliography{IEEEabrv,references}

\end{document}